\documentclass[3p]{elsarticle} 

\myfooter[R]{First version: April 6, 2022. This version: June 15, 2022.}

\usepackage{amscd}      
\usepackage{amsbsy}     
\usepackage{amsmath}    
\usepackage{amsfonts}   
\usepackage{amsthm}     
\usepackage{enumerate}  
\usepackage{fancybox}   
\usepackage{geometry}   

\usepackage[linesnumbered,boxed,longend]{algorithm2e} 

\usepackage{tikz}       
\usetikzlibrary{arrows,calc,decorations,shapes}

\usepackage{cancel}     

\usepackage{caption}    
\usepackage{fancyhdr}   
\usepackage{float}      
\usepackage{graphics}   
\usepackage{graphicx}   
\usepackage{subcaption} 

\usepackage{bbm}        
\usepackage[OT1]{fontenc}

\usepackage{bookmark}   
\usepackage{hyperref}   
\usepackage{cleveref}   

\usepackage{hhline}     
\usepackage{longtable}  
\usepackage{multirow}   
\usepackage{pdflscape}  
\usepackage{tabularx}   

\numberwithin{equation}{section}	                                                    

\DeclareMathOperator{\DF}{DF}

\DeclareMathOperator{\ZC}{ZC}

\DeclareMathOperator{\LGD}{LGD}

\newcommand{\xva}{\text{xVA}}
\newcommand{\xVA}{\xva}

\DeclareMathOperator{\CVA}{CVA}

\DeclareMathOperator{\DVA}{DVA}

\DeclareMathOperator{\FVA}{FVA}
\DeclareMathOperator{\FCA}{FCA}
\DeclareMathOperator{\FBA}{FBA}

\DeclareMathOperator{\EPE}{EPE}

\DeclareMathOperator{\WWR}{WWR}

\newcommand{\EPEFVA}[2]{\EPE_{\FVA}(#1;#2)}
\newcommand{\EPEFVAWWR}[2]{\EPE_{\FVA}^{\WWR}(#1;#2)}
\newcommand{\EPEFVAIndep}[2]{\EPE_{\FVA}^{\perp}(#1;#2)}
\newcommand{\FVAWWR}{\FVA^{\WWR}}
\newcommand{\FVAIndep}{\FVA^{\perp}}

\DeclareMathOperator{\PnL}{P\&L}

\renewcommand{\d}{{\rm d}}
\newcommand{\e}{{\rm e}}                
\newcommand{\E}{\mathbb{ E}}            
\newcommand{\F}{\mathcal{F}}            
\newcommand{\G}{\mathcal{G}}            
\renewcommand{\H}{\mathcal{H}}          
\newcommand{\I}{\mathbbm{1}}            

\newcommand{\Q}{\mathbb{ Q}}            

\def\dt{{\d}t}
\def\du{{\d}u}
\def\dv{{\d}v}

\def\dx{{\d}x}

\newcommand{\maxOperator}[1]{\left( #1 \right)^+}

\newcommand{\indicator}[1]{\I_{\left\{#1\right\}}}

\newcommand{\expPower}[1]{\e^{#1}}

\newcommand{\rdef}{=:}

\newcommand{\cov}{\mathbb{C}\text{ov}}      

\newcommand{\condExp}[2]{\E\left[ \left. #1 \right| \F(#2)\right]}

\newcommand{\condExpSmall}[2]{\E_{#2}\left[ #1 \right]}

\newcommand{\condCovSmall}[3]{\cov_{#3}\left(#1,#2\right)}

\newcommand{\zeroRomanUpperCase}[1]{\ifcase #1 \relax 0 \else {\MakeUppercase{\romannumeral #1}}\fi}

\newtheorem{lem}{Lemma}

\newtheorem{assumption}{Assumption}

\newtheorem*{rem}{Remark}

\newcommand{\resultFigureSize}{0.49\linewidth}

\newcommand{\collateralColor}{gray}

\newcommand{\fixedLegColor}{blue}
\newcommand{\floatLegColor}{red}
\newcommand{\csaColor}{blue!20}
\newcommand{\dealColor}{red!20}

\newcommand{\collateralAccount}{C}

\newcommand{\liquidity}{\ell}

\newcommand{\shortRate}{r}

\newcommand{\borrowingRate}{\shortRate_b}
\newcommand{\zcb}{P}

\newcommand{\spread}{s}

\newcommand{\borrowingSpread}{\spread_b}
\newcommand{\lendingSpread}{\spread_l}

\newcommand{\tradeVal}{V}

\newcommand{\brownian}{W}

\newcommand{\intensity}{\lambda}

\newcommand{\corr}{\rho}
\newcommand{\vol}{\sigma}
\newcommand{\default}{\tau}

\newcommand{\impliedVolFun}[1]{\vol_{\text{imp},#1}} 

\title{Relevance of Wrong-Way Risk in Funding Valuation Adjustments}

\begin{document}

\author[1,2]{Thomas van der Zwaard\corref{cor1}}
\ead{T.vanderZwaard@uu.nl}
\author[1,2]{Lech A.~Grzelak}
\ead{L.A.Grzelak@uu.nl}
\author[1]{Cornelis W.~Oosterlee}
\ead{C.W.Oosterlee@uu.nl}
\cortext[cor1]{Corresponding author at Mathematical Institute, Utrecht University, Utrecht, the Netherlands.}
\address[1]{Mathematical Institute, Utrecht University, Utrecht, the Netherlands}
\address[2]{Rabobank, Utrecht, the Netherlands}

\begin{abstract}
    \noindent In March 2020, the world was thrown into financial distress.
    This manifested itself in increased uncertainty in the financial markets.
    Many interest rates collapsed, and funding spreads surged significantly, which increased due to the market turmoil.
    In light of these events, it is essential to understand and model Wrong-Way Risk (WWR) in a Funding Valuation Adjustment ($\FVA$) context.
    WWR may currently be absent from $\FVA$ calculations in banks’ Valuation Adjustment ($\xVA$) engines.
    However, in this letter, we demonstrate that WWR effects are non-negligible in $\FVA$ modelling from a risk-management perspective.
    We look at the impact of various modelling choices, such as including the default times of the relevant parties, as well as stochastic and deterministic funding spreads.
    A case study is presented for interest rate derivatives.
\end{abstract}

\begin{keyword}
     Wrong-Way Risk (WWR) \sep Funding Valuation Adjustment ($\FVA$) \sep computational finance \sep risk management
    \JEL C63 \sep G01 \sep G13 \sep G32
\end{keyword}

\maketitle

{\let\thefootnote\relax\footnotetext{The views expressed in this paper are the personal views of the authors and do not necessarily reflect the views or policies of their current or past employers. The authors have no competing interests.}}

\section{Introduction}  \label{sec:introduction}

Suppose a corporate has a loan from a bank.
Typically, the cheaper loans are based on a floating rate, paying a variable interest rate (IR), e.g., a Libor rate.
When rates go up, the company has increased costs.
To hedge against this, a company often purchases a payer IR swap (payer means that the company will pay the fixed rate and receive float) from a bank.
From the perspective of the bank, this is a receiver swap.
This way, the company has hedged the floating IR risk and only pays a fixed rate.
Since the corporate does not post any security (collateral), this is an uncollateralized trade.

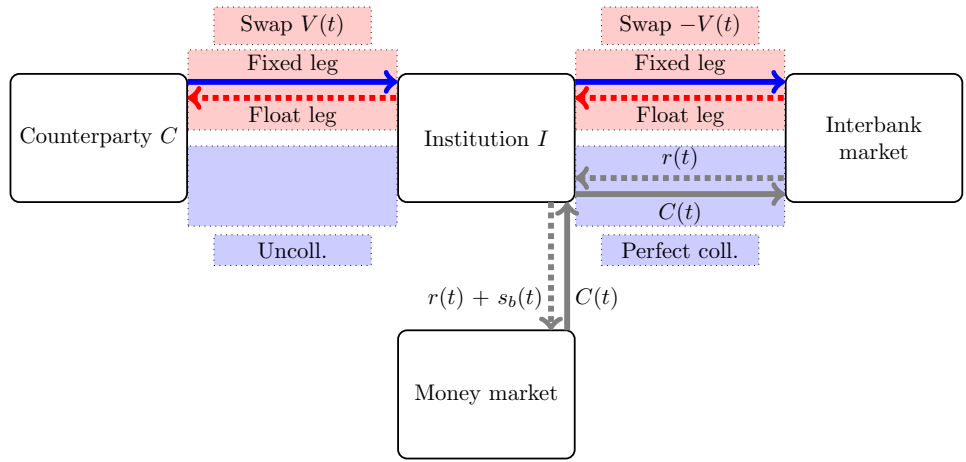
\begin{figure}[h]
    \centering
    \begin{subfigure}[b]{0.85\linewidth}
    \scalebox{0.85}{
    \begin{tikzpicture}
        \node[rectangle, draw, thick, text width=2.5cm, text centered, rounded corners, minimum height=2.0cm] (Counterparty) at (-2,-2) {Counterparty $C$};
        \node[rectangle, draw, thick, text width=2.5cm, text centered, rounded corners, minimum height=2.0cm] (Bank) at (4,-2) {Institution $I$};
        \node[rectangle, draw, thick, text width=2.5cm, text centered, rounded corners, minimum height=2.0cm] (InterbankMarket) at (10,-2) {Interbank market};
        \node[rectangle, draw, thick, text width=2.5cm, text centered, rounded corners, minimum height=2.0cm] (MoneyMarket) at (4,-6) {Money market};

        \node[rectangle, draw, dotted, text width=3.0cm, text centered, minimum height=1.25cm, fill=\csaColor] (NoCSA) at (1,-2.75) {};
        \node[rectangle, draw, dotted, text width=2.2cm, text centered, minimum height=0.4cm, fill=\csaColor] (NoCSALabel) at (1,-3.75) {Uncoll.};
        \node[rectangle, draw, dotted, text width=3.0cm, text centered, minimum height=1.25cm, fill=\csaColor] (CSA) at (7,-2.75) {};
        \node[rectangle, draw, dotted, text width=2.2cm, text centered, minimum height=0.4cm, fill=\csaColor] (CSALabel) at (7,-3.75) {Perfect coll.};

        \node[rectangle, draw, dotted, text width=3.0cm, text centered, minimum height=1.25cm, fill=\dealColor] (Trade) at (1,-1.25) {};
        \node[rectangle, draw, dotted, text width=2.2cm, text centered, minimum height=0.4cm, fill=\dealColor] (TradeLabel) at (1,-0.25) {Swap $\tradeVal(t)$};
        \node[rectangle, draw, dotted, text width=3.0cm, text centered, minimum height=1.25cm, fill=\dealColor] (Hedge) at (7,-1.25) {};
        \node[rectangle, draw, dotted, text width=2.2cm, text centered, minimum height=0.4cm, fill=\dealColor] (HedgeLabel) at (7,-0.25) {Swap $-\tradeVal(t)$};


        \draw[->, line width=2.6pt, color=\fixedLegColor] (-0.625,-1.125) -- (2.625,-1.125);
        \node[align=center, above] (CorporateToBank) at (1.0,-1.125) {Fixed leg};
        \draw[<-, line width=2.6pt, dotted, color=\floatLegColor] (-0.625,-1.375) -- (2.625,-1.375);
        \node[align=center, below] (BankToCorporate) at (1.0,-1.375) {Float leg};

        \draw[<-, line width=2.6pt, dotted, color=\floatLegColor] (5.375,-1.375) --  (8.625,-1.375);
        \node[align=center, below] (InterbankMarketToBankColl) at (7.0,-1.375) {Float leg};
        \draw[->, line width=2.6pt, color=\fixedLegColor] (5.375,-1.125) --  (8.625,-1.125);
        \node[align=center, above] (BankToInterbankMarketColl) at (7.0,-1.125) {Fixed leg};


        \draw[<-, line width=2.6pt, dotted, color=\collateralColor] (5.375,-2.625) --  (8.625,-2.625);
        \node[align=center, above] (InterbankMarketToBankSwap) at (7.0,-2.625) {$\shortRate(t)$};
        \draw[->, line width=2.6pt, color=\collateralColor] (5.375,-2.875) -- (8.625,-2.875);
        \node[align=center, below] (BankToInterbankMarketSwap) at (7.0,-2.875) {$\collateralAccount(t)$};

        \draw[<-, line width=2.6pt, color=\collateralColor] (5.25,-3) -- (5.25,-5);
        \node[align=center, right] (MoneyMarketToBank) at (5.25,-4.5) {$\collateralAccount(t)$};
        \draw[->, line width=2.6pt, dotted, color=\collateralColor] (5,-3) -- (5,-5);
        \node[align=center, left] (BankToMoneyMarket) at (5,-4.5) {$\shortRate(t)$ + $\borrowingSpread(t)$};
    \end{tikzpicture}
    }
    \end{subfigure}
    \caption{An uncollateralized swap with value $\tradeVal(t) > 0$, between counterparty $C$ (a corporate) and institution $I$ (a bank).
    The swap consists of a fixed and a floating leg with fixed and variable cash flows, respectively.
    The opposite hedge with value $- \tradeVal(t) < 0$, in the interbank market, with perfect collateralization (coll.).
    All values are denoted from the institution's perspective.
    $I$ needs to post collateral $\collateralAccount(t)$, for the hedge, to the interbank market counterparty.
    The collateral accrues at the risk-free rate $\shortRate(t)$.
    $I$ needs to fund itself in the money market at the cost of a funding spread $\borrowingSpread(t)$ over $\shortRate(t)$.
    Dotted lines refer to variable cash flows, while solid lines refer to fixed cash flows.}
    \label{fig:fundingCostsSchematically}
\end{figure}

On the other hand, the bank now has a swap, which it hedges in the interbank/cleared market, where contracts are typically collateralized.
Hence, the bank has to post collateral to the interbank counterparty, while not receiving any collateral from the corporate.
The bank needs to fund the collateral amount, where it pays a funding spread over the risk-free rate.
This is a funding cost for the bank, which should be included into the swap pricing.
See Figure~\ref{fig:fundingCostsSchematically} for a graphical overview of the situation.

Funding costs of unsecured transactions are incorporated in financial derivatives pricing through the so-called Funding Valuation Adjustment ($\FVA$), a type of valuation adjustment ($\xVA$).
These adjustments to a derivative value reflect credit risk, funding, regulatory capital and margin, see for example,~\cite{BrigoFrancischelloPallavicini201904,BrigoLiuPallaviciniSloth201612,Green201511,Gregory202007}.
$\FVA$ represents the funding cost of eliminating market risk on non-perfectly collateralized deals.
Credit Valuation Adjustment ($\CVA$) is the adjustment for counterparty credit risk.
$\CVA$ and $\FVA$ can together be interpreted as the cost of imperfect collateralization.
For other intuition on $\CVA$ and the hedging thereof, see~\cite{ZwaardGrzelakOosterlee202102}.

\begin{figure}[H]
  \centering
  \includegraphics[scale=0.4]{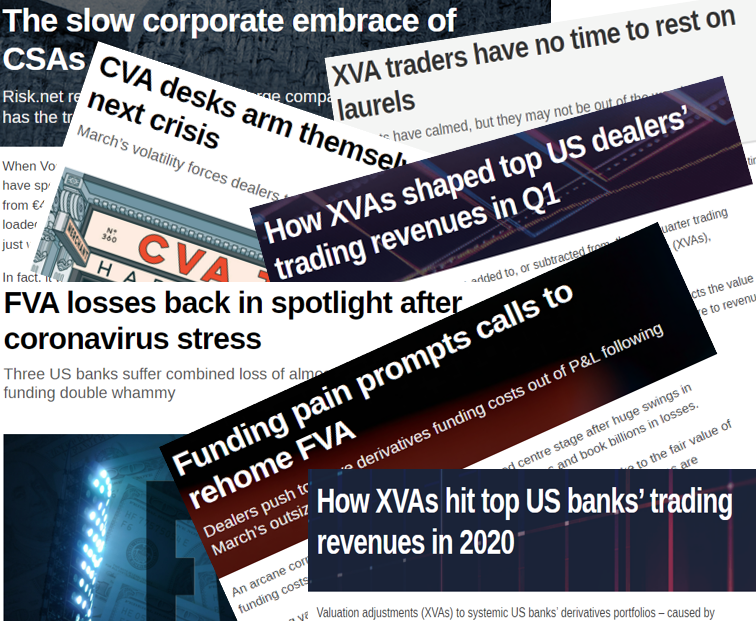}
  \caption{Risk.net headlines~\cite{Risk20200416,Risk20200903,Risk20200824,Risk20201105,Risk20210121,Risk20210309,Risk20200611}.}
  \label{fig:newsHeadlines}
\end{figure}

During the period of financial distress following March 2020, significant market moves took place.
Specifically, interest rates dropped, and funding spreads increased drastically.
If a bank has many corporate clients entering this same type of swap, the bank's portfolio is unbalanced.
When interest rates drop (e.g., due to the central bank's interventions), these swaps move deeply into the money (ITM).~\footnote{ATM (at-the-money) means the current value of the swap is zero. ITM (in-the-money) and OTM (out-of-the-money) indicate that the swap value is respectively positive and negative.}
As a result, the bank needs to post more collateral on the hedge in the interbank market, while not receiving any collateral from the corporate.
The funding requirement on this collateral, combined with exploding funding spreads, could explain the significant losses banks reported following the March 2020 events~\cite{Risk20200416}.
The loss sizes depend on the bank's portfolio composition, valuation methods, counterparty creditworthiness and the bank funding risk.
The market turmoil and corresponding losses had a significant impact on the derivatives business, see Figure~\ref{fig:newsHeadlines}.

According to ISDA~\footnote{International Swaps and Derivatives Association.}, WWR  occurs when ``exposure to a counterparty is adversely correlated with the credit quality of that counterparty''~\cite{ISDA200109}.
This is generic WWR, as opposed to specific, which involves the specifics of a deal structure.
Right-Way Risk (RWR) is the opposite of WWR and occurs when there is a favourable rather than adverse correlation.~\footnote{We will also use the term WWR to indicate RWR, as the difference is only in sign.}

$\FVA$ WWR means increased funding risk due to increased market risk.
For an unbalanced portfolio of receiver swaps, WWR occurs for a negative correlation between interest rates and funding spreads: if IR goes down, exposure goes up, implying that $\FVA$ goes up, which increases the funding spread sensitivity and vice versa.
In addition, the funding spread will go up due to the adverse relationship between IR and funding spread.

\begin{figure}[h]
  \centering
  \includegraphics[scale=0.5]{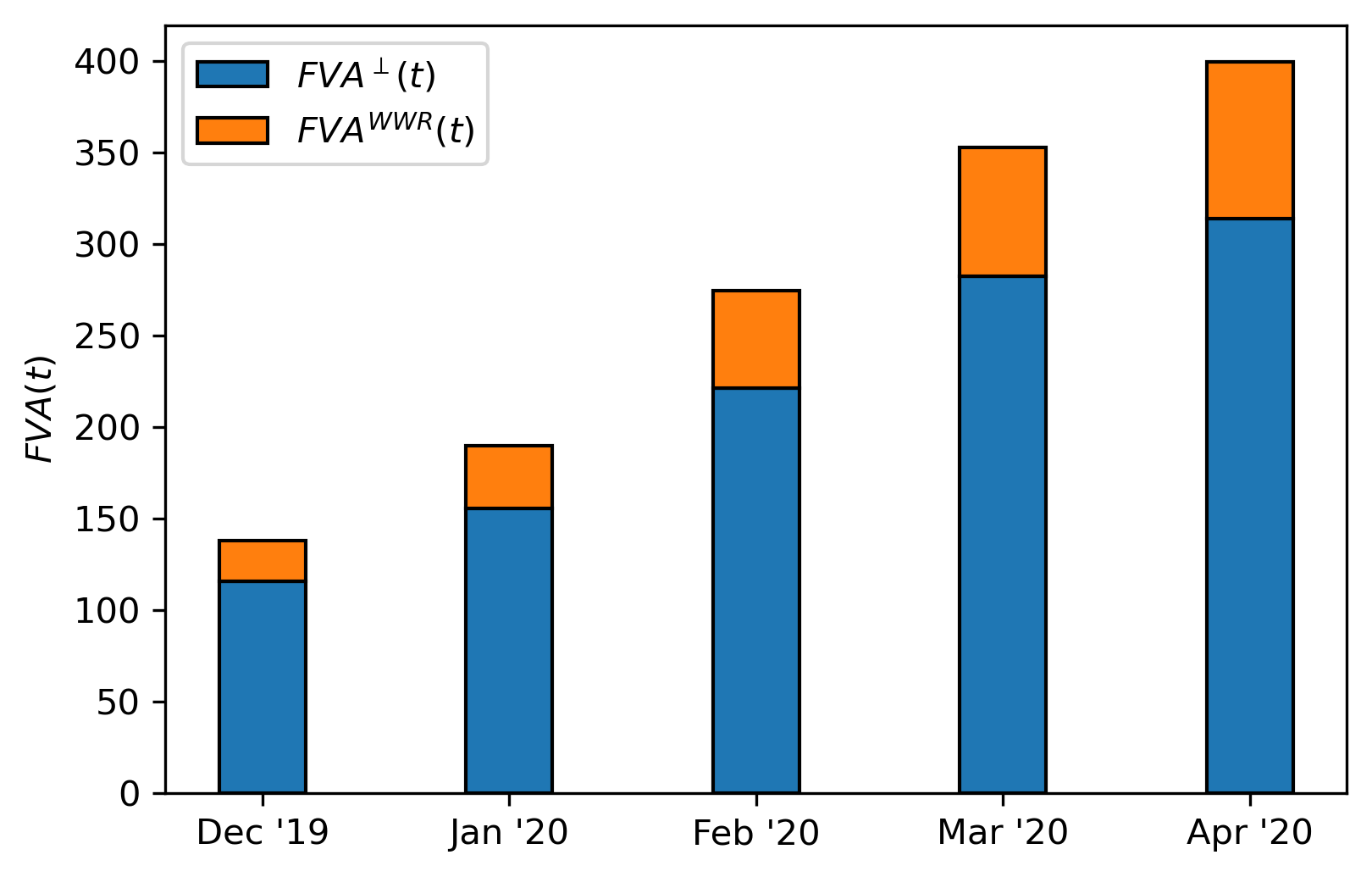}
  \caption{
    $\FVA$ through time for a receiver swap which is ATM in December 2019, with partially synthetic market data.
    There is a split between the independent part, $\FVAIndep$, and the WWR part, $\FVAWWR$.
    Interest rates are negative, and decrease through time, such that the swap becomes ITM.
    The credit spreads are increasing through time.
    The other parameters are handpicked, such that the implied IR and credit volatilities are increasing through time.
    The correlation parameters are kept constant.
    The $\FVA$ results are for a stochastic funding spread, and both party's default times are excluded from the $\FVA$ definition.
    These concepts will be introduced in Section~\ref{sec:wwr}.
  }
  \label{fig:FVAThroughTimeITM}
\end{figure}

To demonstrate the relevance of incorporating WWR in $\FVA$ modelling, historical $\FVA$ is plotted through time in Figure~\ref{fig:FVAThroughTimeITM}.
These results indicate that $\FVA$ can increase significantly under unfavourable market moves.
This is inline with the example from Turlakov~\cite{Turlakov201303}, who demonstrated that the FVA WWR effect can be substantial when considering tail risk.
However, even without WWR, this $\FVA$ increase is significant.

Nevertheless, WWR is non-negligible in $\FVA$ modelling.
$\FVA$ WWR models cross-gamma~\footnote{Cross-gamma risks are second-order partial derivatives w.r.t. two different linear market data inputs.} risks between funding spreads and market exposure.
These cross-gammas are challenging to hedge directly using standard financial derivatives.
Alternatively, the hedging positions can be rebalanced more frequently.
Yet, under stressed market conditions, these hedges become increasingly expensive due to low liquidity.~\footnote{Low liquidity means it is difficult to quickly buy/sell an asset in the market at a price which reflects its current value.}
Hence, the WWR premium can be interpreted as a compensation for increased hedging costs.
Furthermore, the cross-gammas help an $\xVA$ desk to assess their sensitivity to WWR market scenarios.
This helps in risk management, where changing sensitivities can be anticipated when other market factors change.
Rather than waiting for the overnight $\xVA$ calculation to finish and see how sensitivities are affected, the cross-gamma modelling allows the desk to start looking for a suitable hedge on the day the market moves.
The next day, banks with similar books will look for similar hedges, and market liquidity might quickly disappear.
Hence, accurate WWR modelling will help the desk stay within its risk limits.
The effect of adding WWR to the modelling is two-fold: the WWR premium is a compensation for re-hedging at expensive moments, and also recognizing earlier when to re-hedge to limit the hedging costs.
Furthermore, the cross-gammas will help the $\PnL$ explain process~\footnote{The goal of the $\PnL$ explain process is to explain how a portfolio is affected by market movements and other effects (e.g., the passing of time). For more information, see~\cite{ZwaardGrzelakOosterlee202102}.}, to lower the amount of unexplained $\PnL$.

Our contribution in this letter is to understand how various modelling choices affect $\FVA$ WWR.
We focus on including the default times of a trade's parties and the choice of funding spread.
We will see that these choices significantly impact both the $\FVA$ levels and the dependency structure, which is relevant for hedging delta, vega and cross-gamma risks.
The $\FVA$ equation is split into an independent part and a WWR part to assess the WWR effects in isolation.
Including the default times reduces $\FVA$ through a credit adjustment factor.
This factor increases the complexity of the dependency structure.
In our receiver swap examples, this factor even results in RWR, which may seem surprising.
Furthermore, we consider a stochastic funding spread, and remark on the results for a deterministic spread.
The former yields WWR through the stochastic funding spread, possibly dampened by the RWR effects from credit adjustment factors.
The latter results solely in RWR.

\section{FVA and Wrong-Way Risk}  \label{sec:wwr}

There has been a debate in literature on the legitimacy of incorporating $\FVA$ in pricing, initiated by Hull and White~\cite{HullWhite201208}.
Responses came from both academia and practitioners~\cite{BurgardKjaer201210a,Castagna201208,LaughtonVaisbrot201209}, who do not necessarily agree with this.
Hull and White continuously brought new arguments and clarifications to support their statement~\cite{HullWhite201210,HullWhite201405,HullWhite201409,HullWhite201605}.
For an overview of the arguments made in this debate, see~\ref{app:fcaDebate}.
Despite the discussion, market practice has been to include $\FVA$ in pricing.
Yet, the recent funding losses and the difficulty of hedging have some entities advocating to move $\FVA$ out of $\PnL$ statements~\cite{Risk20201105}.

$\FVA$ can be split into a funding benefit ($\FBA$) and cost ($\FCA$).
We assume that no profit can be made on potential funding benefits, i.e., an asymmetric funding assumption.
In particular, a spread over the risk-free rate is paid when borrowing funds, but when lending out, the risk-free rate is earned.
Consequently, $\FBA(t) = 0$, such that $\FVA(t) = \FCA(t)$.
In the case of multiple trades, funding benefits can be present implicitly by reduced funding costs.

We examine $\FVA$ WWR for a single uncollateralized IR derivative $\tradeVal$, between counterparty $C$ and institution $I$, maturing at time $T$.
All values are denoted from $I$'s perspective.
The $\FVA$ is based on borrowing spread $\borrowingSpread >0$ over risk-free rate $\shortRate$.

In Section~\ref{sec:SDE}, we look at the default processes, affine short-rate dynamics used in this work, and go into the correlation structure of the processes.
Then, in Section~\ref{sec:fvaEquation}, we derive the $\FVA$ equation and discuss the impact of including or excluding default times $\default_I$ and $\default_C$ in the $\FVA$ integral.
Furthermore, the $\FVA$ equation and corresponding exposure are split into two parts: an independent part and a WWR part.
In Section~\ref{sec:fundingSpread}, the choice of funding spread is discussed.
This choice is then applied to the $\FVA$ equation in Section~\ref{sec:fvaEquationFundingSpread}, and we end up with an $\FVA$ exposure with WWR.

\subsection{Default processes, model dynamics and correlations} \label{sec:SDE}

We model default times $\default_z$, $z \in \{I,\ C\}$, as the first jumps of a Cox process~\footnote{A Cox process is Poisson process where both the magnitude and the probability of a jump are stochastic~\cite{BrigoMercurio2006}.} with hazard rate (intensity) $\intensity_z$.
We impose affine short-rate models~\cite{OosterleeGrzelak201911} for interest rate $\shortRate$ and hazard rates $\intensity_I$ and $\intensity_C$.
The integrated dynamics are written as:
\begin{align} \displaystyle
  \overline{z}(u)
    &= x_{z}(u) + b_{z}(u), \ \
  x_{z}(u)
    = \mu_{z}(t, u) + y_{z}(t,u), \nonumber
\end{align}
where $\overline{z} \in \{ \shortRate,\ \intensity_I, \ \intensity_C\}$ and subscript $z \in \{\shortRate,\ I,\ C\}$.
Both $b_{z}(u)$ and $\mu_{z}(t, u)$ are deterministic quantities.
Furthermore, $y_{z}(t,u)$ is a stochastic processes, with $\condExpSmall{ y_{z}(t,u)}{t} = 0$.
The affine dynamics imply that the corresponding Zero Coupon Bond, $\zcb_{z}(t,T)$, can be computed analytically.

Dependency between the processes can be introduced by correlating the Brownian motions in $y_{z}(t,u)$~\cite{GarciaMunoz201312} or using a copula~ \cite{BrigoCapponiPallaviciniPapatheodorou201101,BrigoPallaviciniPapatheodorou201107}.
We choose the former, with independent defaults of counterparties $I$ and $C$, which is justifiable as this is not the main driver in WWR modelling.
Since we look at the WWR impact for IR derivatives, the main driver will be the dependency between the funding spread and the IR exposure.~\footnote{When dealing with credit derivatives, this dependency between defaults should definitely be present.}
Furthermore, independence of defaults is not an uncommon assumption in $\xva$ literature, see for example~\cite{BrigoFrancischelloPallavicini201904}.
The independence of defaults allows for factorization of survival probabilities, which turns out to be helpful in the $\FVA$ derivation in Section~\ref{sec:fvaEquation}.
Further motivation for this choice of dependency structure is given in~\ref{app:dependencyStructure}.

In terms of the Brownian motions $\brownian(t)$, the correlation assumptions read
\begin{align} \displaystyle
  \brownian_{\shortRate}(t) \brownian_I(t) = \corr_{\shortRate,I} \cdot t, \
  \brownian_{\shortRate}(t) \brownian_C(t) = \corr_{\shortRate,C} \cdot t, \
  \brownian_I(t) \brownian_C(t) = 0, \nonumber
\end{align}
where the IR-credit correlations $\corr_{\shortRate,I}$ and $\corr_{\shortRate,C}$ can be estimated historically.
If credit data is unavailable, e.g., for illiquid counterparties, techniques exist to map these counterparties to liquid counterparties and the corresponding credit contracts~\cite{Green201511}.

\subsection{FVA equation} \label{sec:fvaEquation}
Starting from the $\FVA$ definition~\cite{AlbaneseAndersenIabichino201502}, we derive the following expression for the $\FVA$ of financial derivative $\tradeVal$ in~\ref{app:fvaDerivation}, under the assumption of independence of defaults.
We assume that no defaults take place before today ($t$).
\begin{align} \displaystyle
  \FVA(t)
    &= \E \left[ \left. \int_t^{T\wedge\default_I\wedge\default_C} \expPower{-\int_{t}^{u} \shortRate(v)\dv} \borrowingSpread(u) \maxOperator{\tradeVal(u)} \du \right| \G(t) \right] \label{eq:fca0} \\
    &=  \int_t^{T} \condExp{ \expPower{-\int_{t}^{u} \intensity_I(v) + \intensity_C(v)\dv}\expPower{-\int_{t}^{u} \shortRate(v)\dv} \borrowingSpread(u) \maxOperator{\tradeVal(u)} }{t} \du\nonumber \\
    &\rdef  \int_t^{T} \EPEFVA{t}{u} \du.  \label{eq:fca1}
\end{align}
Here, $\maxOperator{x} = \max \{x, 0\}$, $\F(t)$ is the `standard' default-free filtration and $\G(t)$ is the enriched filtration with all available market information, including defaults.
Going forward, we write $\condExp{\cdot}{t} = \condExpSmall{\cdot}{t}$.

In Equation~\eqref{eq:fca1}, $\FVA$ represents the cost to fund positive exposure ($\EPE$).~\footnote{
Typically, the $\FVA$ formula is given in terms of a forward funding spread.
However, that is only possible if the funding spread is independent of the exposure, which is currently not the case.}
Hence, $\FVA$ is an integral over the expected valuation profile with the funding spread and accounts for the costs of funding at a different rate than the risk-free rate.
This is a different type of integral than the $\CVA$ integral~\cite{Green201511,Gregory202007}, which is an integral over all possible default times.

In the $\FVA$ definition~\eqref{eq:fca0}, the integration range is $[t,T\wedge\default_I\wedge\default_C]$.~\footnote{This means that we integrate to maturity $T$, or to one of the default times $\default_I$ or $\default_C$, whichever comes first.}
If a party defaults before maturity, $I$ needs to fund for a shorter period than until maturity $T$.
This results in a credit adjustment factors $\expPower{-\int_{t}^{u} \intensity_I(v) \dv}< 1$ and $\expPower{-\int_{t}^{u} \intensity_C(v)\dv} < 1$ for the potential default of $I$ and $C$, which resembles the survival probability of the relevant parties.
It can significantly decrease the overall $\FVA$ amount, depending on the credit quality of the parties.
Hence, the assumption of including $\default_I$ and/or $\default_C$ in the $\FVA$ integral is a crucial modelling choice.

For example, excluding $\default_I$ and $\default_C$ can be regarded as a conservative assumption.~\footnote{Assume for a moment that both $\default_I$ and $\default_C$ are excluded from the $\FVA$ definition, such that in Equation~\eqref{eq:fca0} integration is over the interval $[t,T]$ regardless of potential defaults.
As a result, the credit adjustment factor $\expPower{-\int_{t}^{u} \intensity_I(v) + \intensity_C(v)\dv}$ will disappear from Equation~\eqref{eq:fca1}.
Due to the absence of the credit adjustment factor, the $\FVA$ will be substantially larger in absolute terms compared to the previous formulation.}
This claim is only valid in the asymmetric funding case where no funding benefits are present.

Depending on the correlations, the credit adjustment factors give rise to an extra dependency.
This is particularly interesting if the funding spread is driven by the same underlying source of randomness as a credit adjustment factor, i.e., a party's credit process.
In this situation, the WWR effects can become non-intuitive.

Furthermore, the assumption of including $\default_I$ and/or $\default_C$ in~\eqref{eq:fca0} is also relevant for hedging $\FVA$.
Hence, the modelling assumptions may depend on how an $\xVA$ desk hedges its $\FVA$ risks.
The credit adjustment factor translates into adjusted $\FVA$ sensitivities and introduces new risk-factors to which the $\FVA$ is sensitive.
This impacts first-order delta and vega risks and introduces cross-gamma risks with the existing risk-factors.

\begin{rem}[Additive $\FVA$ assumption]
When $\FVA$ is seen as the funding extension of the $\CVA$-$\DVA$~\footnote{Debit Valuation Adjustment.} paradigm, this will intuitively lead to an extra additive term in the set of $\xva$s.
However, a deal's future cash flows will depend on future funding requirements, hence today's valuation of those cash flows requires the future funding choices to be modelled~\cite{PallaviciniPeriniBrigo201112}.
Hence, $\FVA$ is not an additive $\xva$, but the pricing equation is recursive and highly non-linear.
Rather than using the $\FVA$ definition from Equation~\eqref{eq:fca0}, in absence of $\CVA$ and $\DVA$ terms, $\FVA$ can be defined as ``the difference between the price of the deal when all funding costs are set to zero and the price of the deal when funding costs are included''~\cite{PallaviciniPeriniBrigo201112}.
These recursive systems can be solved using least-squares Monte Carlo regression techniques, see for example the Longstaff-Schwartz method~\cite{LongstaffSchwartz200101}.
Under symmetric funding (equal funding and borrowing rates), the recursive nature of the problem disappears, as the funding requirement does not depend on the sign of the exposure~\cite{PallaviciniPeriniBrigo201212}.
In this case, $\FVA$ can be considered as additive, yet this assumption typically does not hold in practice.
In~\cite{BrigoLiuPallaviciniSloth201612}, the recursive $\FVA$ is compared to the additive version through a Non-linearity Valuation Adjustment.
The authors state that this framework is not suitable for implementation within banks, but allows to analyze the impact of linearization.
Financial institutions often make the simplifying assumption of additive $\FVA$, at the risk of some double counting between the various $\xva$s.
This is also the approach we take, in line with current industry practice.
\end{rem}

$\EPEFVA{t}{u}$ from Equation~\eqref{eq:fca1} can be written as the sum of the independent exposure $\EPEFVAIndep{t}{u}$ and a WWR exposure $\EPEFVAWWR{t}{u}$, i.e.,
\begin{align} \displaystyle
  \EPEFVA{t}{u}
    &= \EPEFVAIndep{t}{u} + \EPEFVAWWR{t}{u}, \label{eq:epeSum}
\end{align}
where the precise form of these exposures for a specific funding spread will follow in Section~\ref{sec:fvaEquationFundingSpread}.
Now, $\FVA$ from Equation~\eqref{eq:fca1} can be split into an independent part and a part that captures the cross-dependencies:
\begin{align} \displaystyle
  \FVA(t)
    &= \int_t^{T} \EPEFVAIndep{t}{u}\du + \int_t^{T}\EPEFVAWWR{t}{u} \du
    \rdef \FVAIndep(t) + \FVAWWR(t).  \label{eq:fca2}
\end{align}

\subsection{Funding spread} \label{sec:fundingSpread}
The funding rate should reflect an institution's funding abilities in the market.
We mainly focus on a stochastic funding spread containing institution's credit, $\intensity_I(t)$, and a possible liquidity adjustment term $\liquidity(t)$.
For example, $\intensity_I(t)$ can be CDS-based, and $\liquidity(t)$ can be the bond-CDS basis.~\footnote{The basis is negative when the CDSs spreads are lower than the bond spread for the same maturity. In March 2020 this spread was negative due to low liquidity.}
Alternatively, $\intensity_I(t)$ can be bond-based or derived from asset swaps.
We assume that it is CDS-based.
Loss given default $\LGD_I$ is taken constant, based on market information.
We define borrowing spread $\borrowingSpread(t)$ as~\cite{Green201511}:
\begin{align} \displaystyle
  \borrowingSpread(t)
    &= \LGD_I \intensity_I(t) + \liquidity(t). \nonumber
\end{align}

WWR can be introduced through $\borrowingSpread(t)$, if $\intensity_I(t)$ is stochastic and correlated with the other risk-factors.
Using the model dynamics from Section~\ref{sec:SDE}, the borrowing spread is split into a deterministic component $\mu_{S}(t, u)$ and a stochastic component $y_{I}(t,u)$:
\begin{align} \displaystyle
  \borrowingSpread(u)
    &= \LGD_I\left[x_I(u) + b_I(u)\right] + \liquidity(u) \nonumber \\
    &= \LGD_I \left[\mu_{I}(t, u) + b_I(u)\right] + \liquidity(u) + \LGD_I y_{I}(t,u) \nonumber \\
    &\rdef \mu_{S}(t, u) + \LGD_I y_{I}(t,u). \label{eq:fundingSpreadCredit}
\end{align}

Alternatively, the spread can be purely deterministic.
Then, no WWR is introduced through the funding spread, but through the credit adjustment factors and exposure only.
An example of a deterministic funding spread is one based on a borrowing rate $\borrowingRate(t)$ which is a constant spread $\spread$ over an $n$-month ($n$M) term rate, e.g., the forward-looking Libor/Euribor or backward-looking SOFR/ESTR.
We assume the term rate is a deterministic spread term structure $\spread_{\text{$n$M}}(t)$ over OIS rate $\shortRate(t)$.
This yields a deterministic borrowing spread:
\begin{align} \displaystyle
  \borrowingSpread(t)
    &= \borrowingRate(t) - \shortRate(t) = \spread_{\text{$n$M}}(t) + \spread. \label{eq:fundingSpreadIR}
\end{align}
This deterministic spread is a special case of the stochastic spread: when removing the stochasticity from the latter, i.e., setting $y_I(t,u) = 0$, and choosing the drift $\mu_{S}(t, u)$ appropriately, the two cases collapse.

\subsection{FVA exposure under funding spread assumptions} \label{sec:fvaEquationFundingSpread}
The funding spread assumptions of Section~\ref{sec:fundingSpread} can now be applied to the independent and WWR exposures from Equation~\eqref{eq:epeSum}.
Derivations of the exposures presented here are available in~\ref{app:exposureDerivation}.
$\EPEFVAIndep{t}{u}$ is written as:
\begin{align} \displaystyle
  \EPEFVAIndep{t}{u}
    &= \zcb_{I}(t,u) \zcb_{C}(t,u)\mu_{S}(t, u)\condExpSmall{\expPower{-\int_{t}^{u} \shortRate(v)\dv}\maxOperator{\tradeVal(u)}}{t}  \nonumber \\
    &\quad + \LGD_I\condExpSmall{\expPower{-\int_{t}^{u} \intensity_I(v) + \intensity_C(v)\dv} y_I(t,u)}{t} \condExpSmall{\expPower{-\int_{t}^{u} \shortRate(v)\dv}\maxOperator{\tradeVal(u)}}{t}.\label{eq:epeIndepCredita}
\end{align}
Here, survival probabilities $\zcb_{I}(t,u)$ and $\zcb_{C}(t,u)$ are independent, resulting from the correlation assumptions in Section~\ref{sec:SDE}.
The $\mu_{S}$-term in~\eqref{eq:epeIndepCredita} matches the classical case of exposure without WWR.
The $y_I$-term captures the dependency between the borrowing spread and the credit adjustment factors.

Using the correlation assumptions from Section~\ref{sec:SDE}, WWR will be present:
\begin{align} \displaystyle
  \EPEFVAWWR{t}{u}
    &= \condExpSmall{\left(\expPower{-\int_{t}^{u} \shortRate(v)\dv}\maxOperator{\tradeVal(u)} - \condExpSmall{\expPower{-\int_{t}^{u} \shortRate(v)\dv}\maxOperator{\tradeVal(u)}}{t} \right)\expPower{-\int_{t}^{u} \intensity_I(v) + \intensity_C(v)\dv}\borrowingSpread(u) }{t}, \label{eq:epeWWRCredit1}
\end{align}
and $\EPEFVAWWR{t}{u}=0$ if IR and credit are independent.

The exposures $\EPEFVAIndep{t}{u}$ and $\EPEFVAWWR{t}{u}$ can now be used to compute $\FVAIndep(t)$ and $\FVAWWR(t)$ in Equation~\eqref{eq:fca2}.
This is done in Section~\ref{sec:wwrRelevance} to examine the WWR effects and the impact of various modelling choices.

\begin{rem}[Alternative $\FVA$ definition]
If $\default_I$ and $\default_C$ are excluded from the $\FVA$ definition,  which is an assumption sometimes made in practice~\cite{Gregory202007}, Equations~(\ref{eq:epeIndepCredita}--\ref{eq:epeWWRCredit1}) change into
\begin{align} \displaystyle
  \EPEFVAIndep{t}{u}
    &= \mu_{S}(t, u)\condExpSmall{\expPower{-\int_{t}^{u} \shortRate(v)\dv}\maxOperator{\tradeVal(u)}}{t}, \label{eq:epeIndepCreditAlt} \\
  \EPEFVAWWR{t}{u}
    &=  \LGD_I \condExpSmall{\expPower{-\int_{t}^{u} \shortRate(v)\dv}\maxOperator{\tradeVal(u)} y_I(t,u)}{t}. \label{eq:epeWWRCredit1Alt}
\end{align}
Here, the credit adjustment factors have disappeared and a term has vanished, resulting in a more simplistic model.
\end{rem}

\begin{rem}[Deterministic funding spread]
For a deterministic funding spread, Equations~(\ref{eq:epeIndepCredita}--\ref{eq:epeWWRCredit1}) change into:
\begin{align} \displaystyle
  \EPEFVAIndep{t}{u}
    &= \zcb_{I}(t,u) \zcb_{C}(t,u)\mu_{S}(t, u)\condExpSmall{\expPower{-\int_{t}^{u} \shortRate(v)\dv}\maxOperator{\tradeVal(u)}}{t},\label{eq:epeIndepCreditaIR} \\
  \EPEFVAWWR{t}{u}
    &=  \borrowingSpread(u)\condExpSmall{\left(\expPower{-\int_{t}^{u} \shortRate(v)\dv}\maxOperator{\tradeVal(u)} - \condExpSmall{\expPower{-\int_{t}^{u} \shortRate(v)\dv}\maxOperator{\tradeVal(u)}}{t} \right)\expPower{-\int_{t}^{u} \intensity_I(v) + \intensity_C(v)\dv} }{t}. \label{eq:epeWWRCredit1IR}
\end{align}

So for $\EPEFVAIndep{t}{u}$ from Equation~\eqref{eq:epeIndepCredita}, the second term drops out.
In Equation~\eqref{eq:epeWWRCredit1}, $\borrowingSpread(u)$ can be moved outside the expectation, and only the correlation between the exposure and the credit adjustment factors remains.
\end{rem}

\section{FVA Wrong-Way Risk relevance} \label{sec:wwrRelevance}

We illustrate the relevance of including WWR in $\FVA$ modelling using numerical examples.
Furthermore, we give insights on the inclusion of $\default_I$ and/or $\default_C$ in the $\FVA$ definition.
Finally, we consider the stochastic and deterministic funding spreads, and show the different WWR/RWR effects in both cases.
We assess the correlation impact on $\FVA$ through the ratio $\frac{\FVA(t)}{\FVAIndep(t)}$.
A ratio larger than 1 corresponds to WWR, while a ratio smaller than 1 corresponds to RWR.

We use the $\FVA$ definition from Equation~\eqref{eq:fca2}.
Using the funding spread assumption of Section~\ref{sec:fundingSpread}, independent exposures $\EPEFVAIndep{t}{u}$ and WWR exposures $\EPEFVAWWR{t}{u}$ are further specified in respectively Equations~\eqref{eq:epeIndepCredita} and~\eqref{eq:epeWWRCredit1}.
We consider a 30-year receiver swap with $10000$ notional.~\footnote{Such that all results can be interpreted as basis point effects.}
These results can easily be extended to other financial derivatives.
$\FVA$ is computed using a Monte Carlo simulation with $10^5$ paths and $10$ dates per year.

In Section~\ref{sec:dynamicsAndCalibration}, we give the affine IR and credit dynamics that fit into the framework introduced in Section~\ref{sec:SDE}, and we discuss the model calibration stage.
In Section~\ref{sec:marketData}, the choice of market data is motivated.
Then we assess $\FVA$ WWR for different modelling assumptions: we consider both the stochastic and deterministic funding spreads in Sections~\ref{sec:wwrRelevanceCreditBasedSpread} and~\ref{sec:wwrRelevanceIRBasedSpread}, respectively, and look at including/excluding $\default_I$ and/or $\default_C$ the $\FVA$ definition.
Furthermore, in Section~\ref{sec:additionalResultsMarketConditions}, the impact of different market conditions and combinations of model parameters is assessed.
There, we also look at payer swaps and the three types of moneyness.

In~\ref{app:additionalResultsModellingAssumptions}, additional results are presented for the various modelling assumptions, where we also present several exposure profiles.
These extra results support the conclusions from the experiments in this section.

\subsection{Model dynamics and calibration} \label{sec:dynamicsAndCalibration}
For IR process $\shortRate(t)$, the Hull-White dynamics with constant volatility are used:
\begin{align} \displaystyle
  \shortRate(t)
    &= x_{\shortRate}(t) + b_{\shortRate}(t), \
  \dx_{\shortRate}(t)
    = -a_{\shortRate}x_{\shortRate}(t)\dt + \vol_{\shortRate} \d\brownian_{\shortRate}(t). \label{eq:hw1f}
\end{align}
The model can be calibrated to a yield curve and a strip of co-terminal swaptions quoted in the market.
For sake of simplicity, we focus on a single swaption implied volatility $\impliedVolFun{\shortRate}$ to which the IR model volatility $\vol_{\shortRate}$ can be calibrated.
Hence, the calibration of the IR process is fully market implied.

A CIR type of model is commonly used in WWR modelling for BCVA purposes~\cite{BrigoCapponiPallaviciniPapatheodorou201101,BrigoPallavicini201405,BrigoPallaviciniPapatheodorou201107}.
Hence, for the credit processes $\intensity_z$, $z \in \{I,\ C\}$, we choose a CIR credit dynamics with constant volatility:
\begin{align} \displaystyle
  \intensity_z(t)
    &= x_z(t) + b_z(t), \
  \dx_z(t)
    = a_z\left( \theta_z -  x_z(t)\right) \dt + \vol_z \sqrt{x_z(t)}\d\brownian_z(t). \label{eq:cir}
\end{align}
For these dynamics, potential issues around the origin might arise.
When the Feller condition $2a_z\theta_z > \vol_z^2$ is satisfied, we are not affected by this issue, see~\cite[p275]{OosterleeGrzelak201911} for more information.

The CIR credit dynamics calibration should be partially market implied and partially historical.
The choice of historical calibration is made due to insufficient liquidity in the CDS option markets.
During the credit calibration, IR and credit are assumed to be independent, such that separability between discount factors and survival probabilities is possible~\cite{BrigoMercurio2006}.
For each institution $z\in\{I,C\}$, the mean reversion $a_z$ and volatility $\vol_z$ parameters can be calibrated to a time series of credit spreads.
For a credit process, the intensity should be positive, i.e., $\intensity_z(t) = x_z(t) + b_z(t) >0$.
Then, $x_z(0)>0$ and $\theta_z$ can be chosen such that the Feller condition is satisfied, as well as the condition $b_z(t) = f^{Market}_z(0,t) - f^{CIR}_z(0,t) > 0$ for all $t$.
In other words, the model implied term structure $f^{CIR}_z$  must fit the market term structure $f^{Market}_z$ from the credit curve as closely as possible, while keeping $b_z(t)>0$.
Given $x_z(0)$, $\theta_z$ can be explicitly computed as a function of the other parameters.~\footnote{In particular, we choose the minimum of the implied $\theta_z$ values for all spine dates, where the implied $\theta_z$ is computed based on $b_z(t) = 0$ for each pillar date $t$.
The positivity constraints leads to $x_z(0) \leq f^M_z(0,0)$.
Ideally we want to minimize $\int_0^t b_z(u)\du$ for all $t$, such that the stochastic part of the intensity captures the maximum of the market quotes.}

\subsection{Market data} \label{sec:marketData}
The market data used for the experiments in this section are those of April 2020 when markets were stressed, following financial distress in March 2020.
In the beginning of 2020, interest rates were negative and continued to drop; the credit curves were relatively stable.
Starting March 2020, credit quality across the spectrum started to deteriorate, resulting in higher market implied default probabilities.
By April 2020, this lower credit quality was still visible, while interest rates were dropping even further.
Clearly, these market conditions match the WWR scenario with an adverse relationship between IR and credit.

Next to market implied data, we use synthetic market data and model parameters.
For the yield curves, we use either a market implied EUR 1D curve or a synthetic $5\%$ flat ZC curve, see~\ref{app:yieldCurves}.
For the credit curves, we consider market implied curves comparable to AAA, BBB or B credit ratings, see~\ref{app:creditCurves}.
Correlation parameters are also synthetic in our experiments, but can also be calibrated historically.

We consider low and high values for mean reversion and model volatility for all dynamics, as well as multiple correlation magnitudes and signs.
Furthermore, we assume that $C$ equal or lower credit quality than $I$: the survival probabilities of $C$ are equal to or lower than those of $I$, and the credit volatility for $C$ is equal to or higher than for $I$.
Without loss of generality, we set liquidity adjustment $\liquidity(t)=0$.
In~\ref{app:scenarios}, scenarios are given on which the experiments are based.

\subsection{Stochastic funding spread} \label{sec:wwrRelevanceCreditBasedSpread}
The results in this section are based on market data scenario~\ref{app:scenario11} in~\ref{app:scenarios}.
The credit adjustment effect of including $\default_I$ and/or $\default_C$ in Equation~\eqref{eq:fca0} is visible from the $\FVAIndep(t)$ values in Table~\ref{tab:fvaCreditBasedSpread}.
This effect is the strongest for $\default_C$, as $C$ has a lower credit quality than $I$.
When including both $\default_I$ and $\default_C$, the combined effects result in the lowest $\FVAIndep(t)$.
The $\FVAIndep(t)$ reduction can be substantial, illustrated by a 74 basis point reduction in this example, which is approximately a $70\%$ decrease.
\begin{table}[h]
    \centering
    \begin{tabular}{l|rr}
                       & $\default_I$ excl. & $\default_I$ incl. \\ \hline
    $\default_C$ excl. & 107.64             & 95.31              \\
    $\default_C$ incl. & 36.10              & 33.63
    \end{tabular}
    \caption{$\FVAIndep(t)$ for the various choices of including/excluding $\default_I$ and/or $\default_C$.}
    \label{tab:fvaCreditBasedSpread}
\end{table}

\begin{figure}[h]
  \centering
  \begin{subfigure}[b]{\resultFigureSize}
    \includegraphics[width=\linewidth]{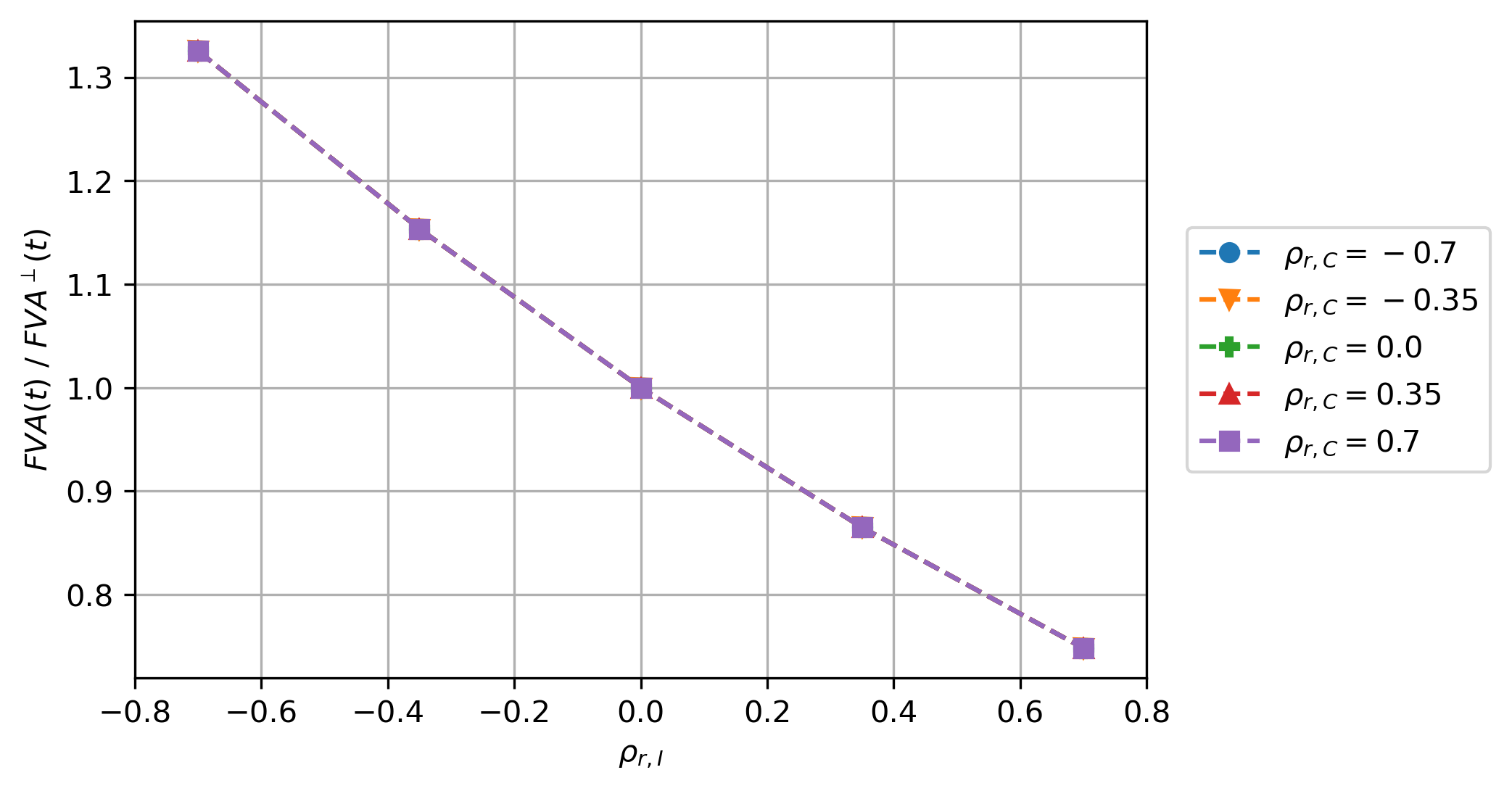}
    \caption{$\default_I$ incl., $\default_C$ excl., $\FVAIndep(t) = 95.31$.}
    \label{fig:corrCreditReceiver20InclExcl}
  \end{subfigure}
  \begin{subfigure}[b]{\resultFigureSize}
    \includegraphics[width=\linewidth]{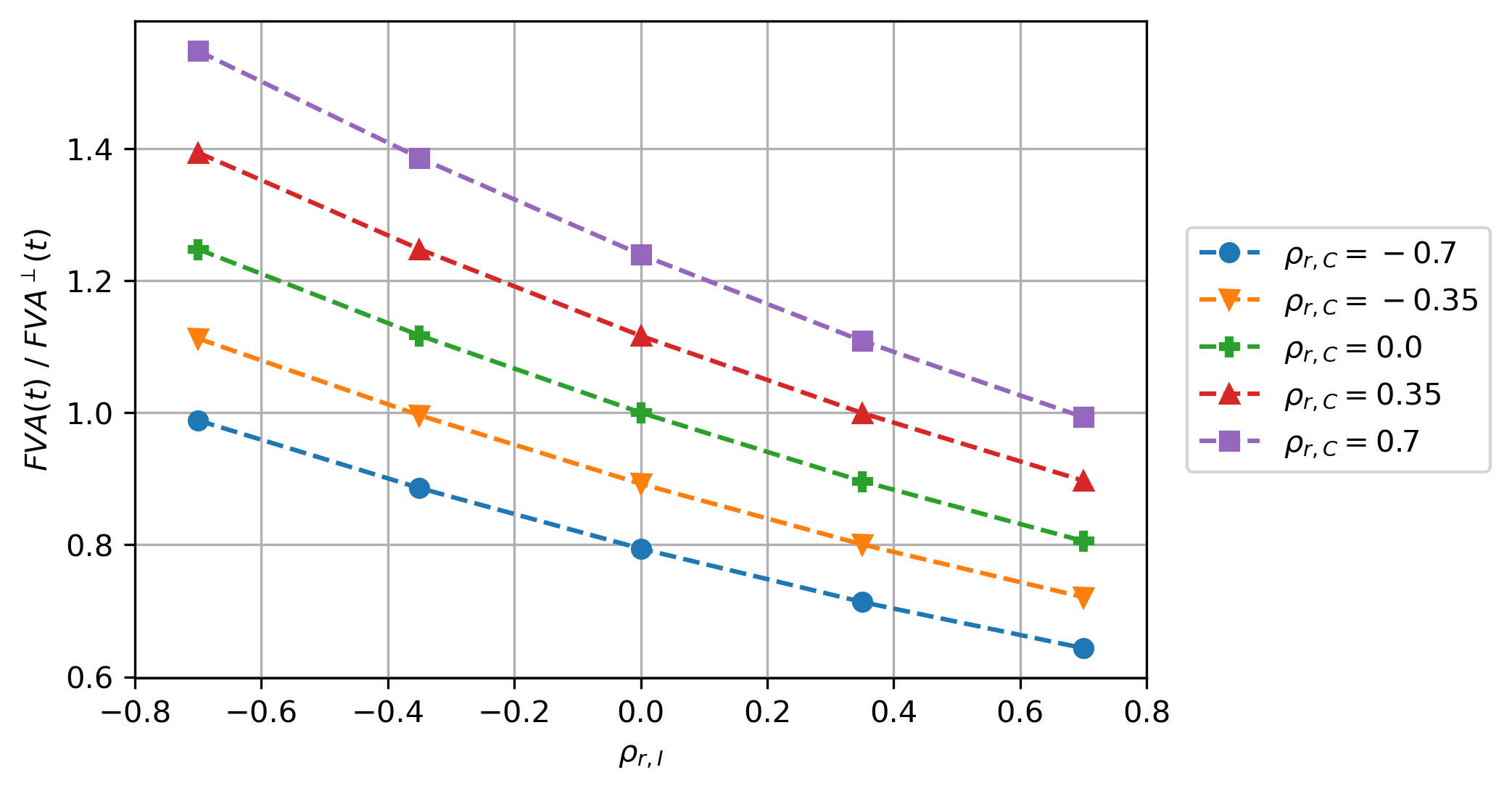}
    \caption{$\default_I$ incl., $\default_C$ incl., $\FVAIndep(t) = 33.63$.}
    \label{fig:corrCreditReceiver20InclIncl}
  \end{subfigure}
  \caption{WWR effects for a stochastic spread and an ATM receiver swap.}
  \label{fig:corrCreditReceiver20}
\end{figure}

In Figure~\ref{fig:corrCreditReceiver20}, the correlation effects are illustrated when including $\default_I$ and changing $\default_C$.
When excluding $\default_I$, similar results are obtained apart from a scaling factor.
When looking at correlation effects, we vary correlations $\corr_{\shortRate,I}$ and $\corr_{\shortRate,C}$ over the interval $[-0.7, 0.7]$.~\footnote{Correlations $\corr_{\shortRate,I}$ and $\corr_{\shortRate,C}$ cannot be varied over the entire interval $[-1.0, 1.0]$ when $\corr_{I,C}=0$, as large absolute IR-credit correlation values then result in a non-SPD correlation matrix.
Hence, $\corr_{\shortRate,I}$ and $\corr_{\shortRate,C}$ are limited to $[-0.7, 0.7]$.
This is not an issue, as extreme correlations lying outside this interval are not required to demonstrate the relevance of WWR.}
The WWR/RWR effects are non-negligible, as ratio $\frac{\FVA(t)}{\FVAIndep(t)}$ is significantly different from 1 for non-zero correlations.

In Figure~\ref{fig:corrCreditReceiver20InclExcl}, there is net WWR for $\corr_{\shortRate,I} < 0$, and the curves for different $\corr_{\shortRate,C}$ values overlap, as $\default_C$ is excluded.
WWR comes from the relationship between the funding spread and the discounted exposure.
This matches the March 2020 market moves: for negative IR-credit correlation, we expect to see WWR for receiver swaps, i.e., the $\FVA$ goes up.
Symmetrically, when the correlation sign flips, i.e., $\corr_{\shortRate,I} > 0$, WWR changes into RWR.
Going forward, we focus on negative correlations, as the symmetry remains: a change in correlation sign changes the WWR/RWR effect.
Furthermore, when including $\default_I$, the credit adjustment effect results in a slight RWR effect for $\corr_{\shortRate,I} < 0$: when excluding $\default_I$, $\frac{\FVA(t)}{\FVAIndep(t)}$ is lower.

For $\corr_{\shortRate,C} < 0$, there is a RWR effect from $C$'s credit adjustment factor: $\frac{\FVA(t)}{\FVAIndep(t)}$ is lower for decreasing $\corr_{\shortRate,C}$, which is apparent when comparing Figures~\ref{fig:corrCreditReceiver20InclExcl} and~\ref{fig:corrCreditReceiver20InclIncl}.
Figure~\ref{fig:corrCreditReceiver20InclIncl} illustrates the effects when including all model components.
Whether there is net WWR or RWR depends on the magnitude of WWR from the funding spread and the degree of the RWR effects from the credit adjustment factors.
In turn, this is driven by the correlation parameters, credit parameters, IR parameters and product type.
In some cases, the RWR effect from the credit adjustment factors outweighs the WWR effect from the funding spread.
For example, when $\corr_{\shortRate,C} = -0.7$, for all values of $\corr_{\shortRate,I}$ there is overall RWR as $\frac{\FVA(t)}{\FVAIndep(t)} < 1$.

Furthermore, when setting one of the two correlations parameters to zero, the correlation magnitude effect is roughly linear in WWR/RWR.
When both correlations are non-zero, the correlation effects can be non-trivial due to the mixing of effects.

Valsecchi comes to similar conclusions on the relevance and linear nature of WWR in the case of an uncollateralized IR swap using a idiosyncratic/systemic decomposition of the stochastic funding spread, where a copula is used for the dependency structure~\cite{Valsecchi202104}.
Yet, in this case, the default times of both parties are excluded from the $\FVA$ definition.

\begin{rem}[Right-Way Risk]
Like WWR, RWR is also a cross-gamma risk, but with an opposite sign.
This sign depends on the correlations, product type and modelling assumptions.
In our examples, RWR comes from a different source in the modelling (the credit adjustment factors) than the WWR (the funding spread).
The type of risk management for RWR is the same as for WWR, as there is only a difference in sign.
In our case, RWR simplifies risk management due to the reduced overall cross-gamma risk.
For an $\xva$ desk, this could imply that hedging positions need to be rebalanced less frequently.
\end{rem}

\begin{rem}[Comparison with $\CVA$]
$\CVA$ WWR/RWR results from the correlation between counterparty default probabilities and exposure.
Like for $\FVA$, WWR increases $\CVA$ and RWR decreases $\CVA$.
However, $\FVA$ depends on $I$'s credit, while $\CVA$ depends on $C$'s credit.
In particular, for $\FVA$, the funding spread $\borrowingSpread = \LGD_I \intensity_I + \liquidity$ appears, which is based on $I$'s credit, whereas for $\CVA$ the factor $\LGD_C \intensity_C$ appears, which is based on $C$'s credit.

Whether the WWR/RWR effect is larger for $\FVA$ or $\CVA$ depends on the differences in the counterparties' credit quality and their correlations with the market risks.
For the following argument, w.l.o.g., consider a receiver IR swap and $\corr_{\shortRate,I},\corr_{\shortRate,C} < 0$.
In the $\FVA$ model, $I$'s credit can generate WWR through borrowing spread $\borrowingSpread$, and RWR through the credit adjustment factor resulting from the inclusion of $\default_I$ in the $\FVA$ formula.
On the other hand, $C$'s credit can generate only RWR through the credit adjustment factor resulting from the inclusion of $\default_C$ in the $\FVA$ formula.
For the $\CVA$ case, this is reversed: $C$'s credit can generate both WWR and RWR, while $I$'s credit can generate only RWR.

Special care is required when also including a Debit Valuation Adjustment, to avoid the double counting of a funding benefit~\cite{Gregory202007}.
In our setup with asymmetric funding, we have $\FBA=0$, so we can also safely include a $\DVA$ component without risking the double counting of a funding benefit.
For $\DVA$, the factor $\LGD_I \intensity_I$ appears.
If $\DVA$ is included, combined with a symmetric funding assumption, i.e., $\borrowingSpread = \lendingSpread = \LGD_I \intensity_I + \liquidity$, the $\LGD_I \intensity_I$ part gets double counted if both $\DVA$ and $\FBA$ are included in the valuation.
\end{rem}

\subsection{Deterministic funding spread} \label{sec:wwrRelevanceIRBasedSpread}
The results in this section are based on market data scenario~\ref{app:scenario11} in~\ref{app:scenarios}.
The credit adjustment effect on $\FVAIndep(t)$ is similar to the stochastic spread case from Table~\ref{tab:fvaCreditBasedSpread}, as $\FVAIndep(t)$ does not depend on the WWR assumptions.~\footnote{Recall that in the deterministic spread case we choose $\borrowingSpread(u) = \mu_S(t,u)$ such that on average the funding spreads are equivalent for deterministic and stochastic funding spreads.}

\begin{figure}[h]
  \centering
  \begin{subfigure}[b]{\resultFigureSize}
    \includegraphics[width=\linewidth]{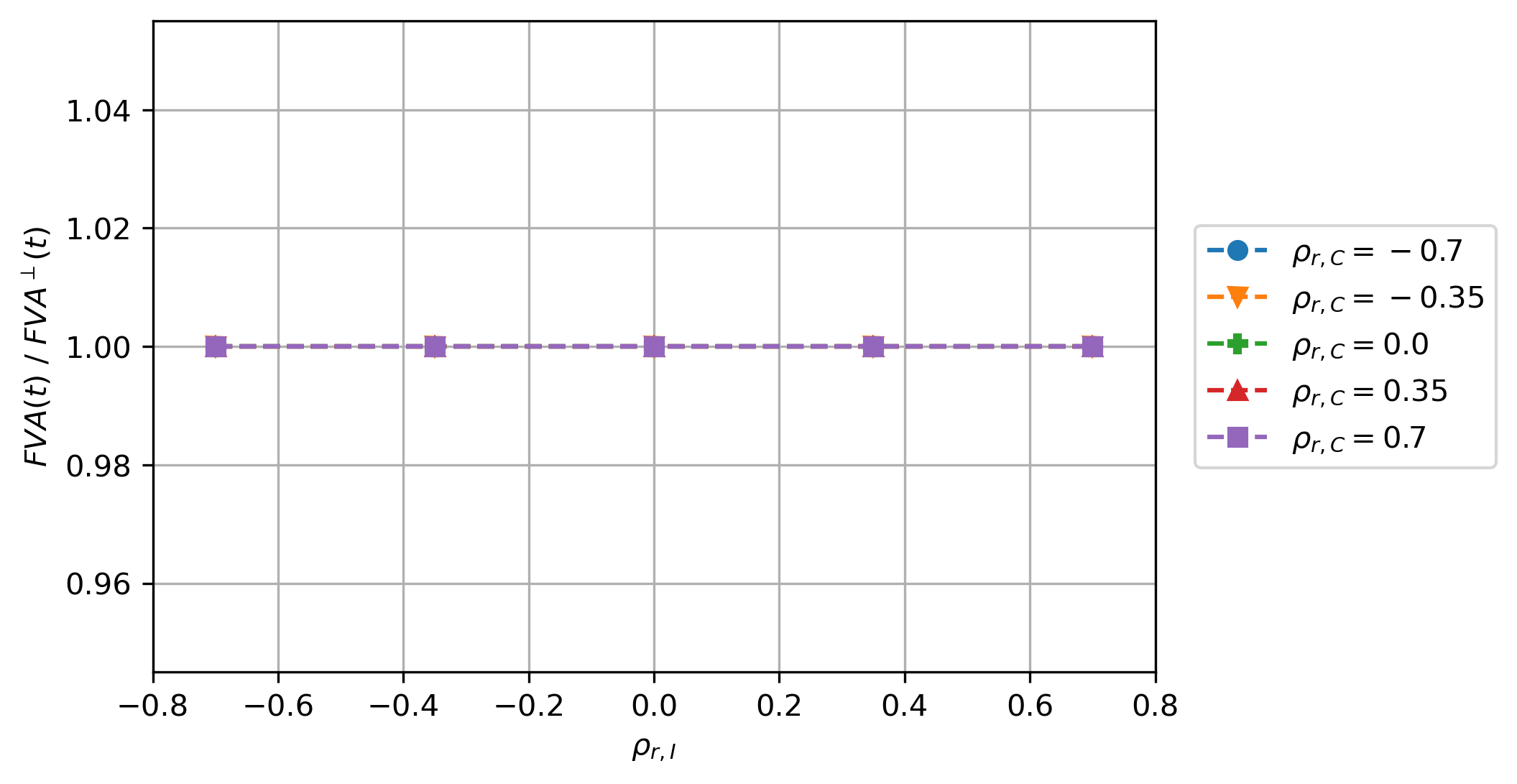}
    \caption{$\default_I$ excl., $\default_C$ excl., $\FVAIndep(t) = 107.63$.}
    \label{fig:corrIRReceiver20ExclExcl}
  \end{subfigure}
  \begin{subfigure}[b]{\resultFigureSize}
    \includegraphics[width=\linewidth]{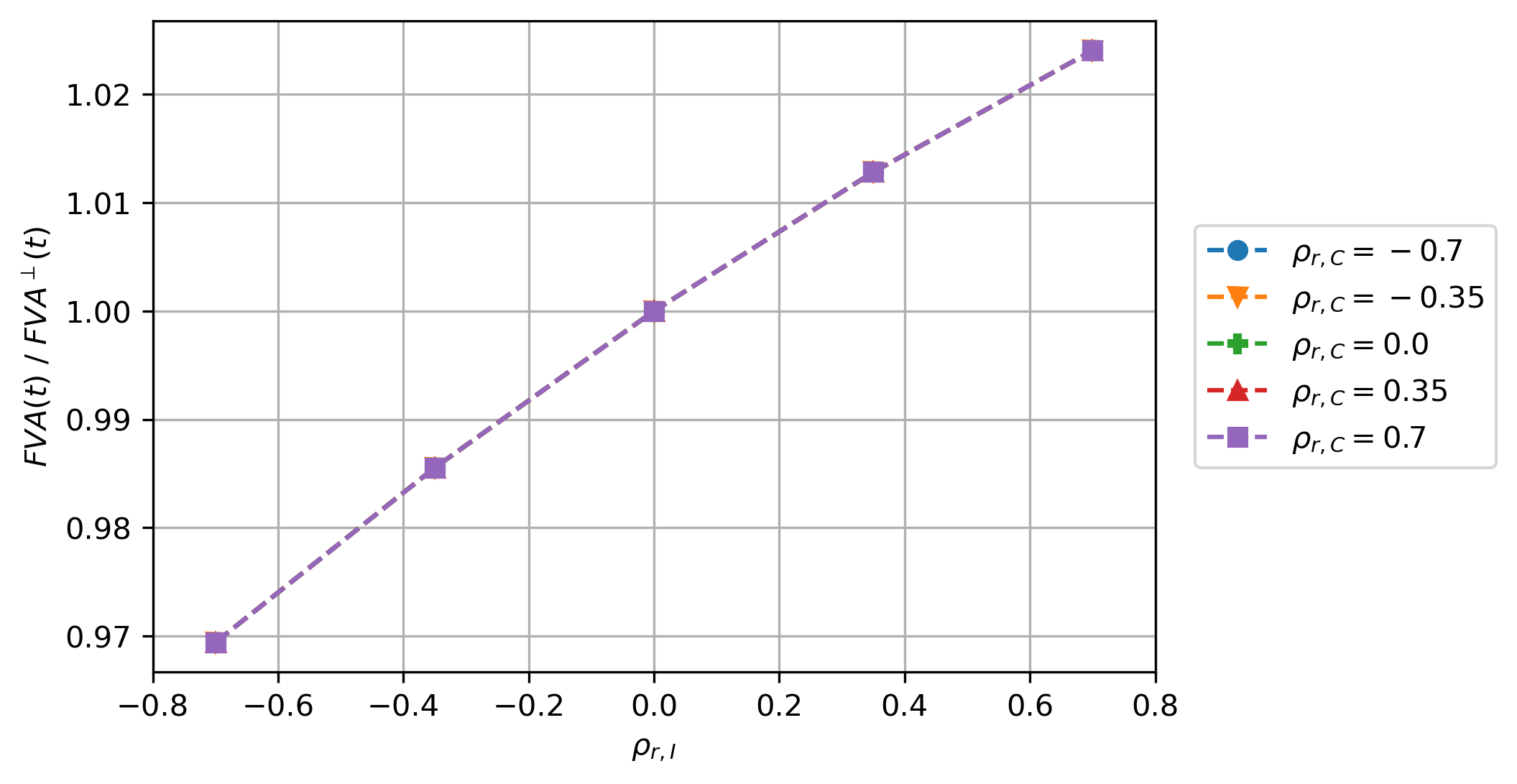}
    \caption{$\default_I$ incl., $\default_C$ excl., $\FVAIndep(t) = 96.19$.}
    \label{fig:corrIRReceiver20InclExcl}
  \end{subfigure}
  \begin{subfigure}[b]{\resultFigureSize}
    \includegraphics[width=\linewidth]{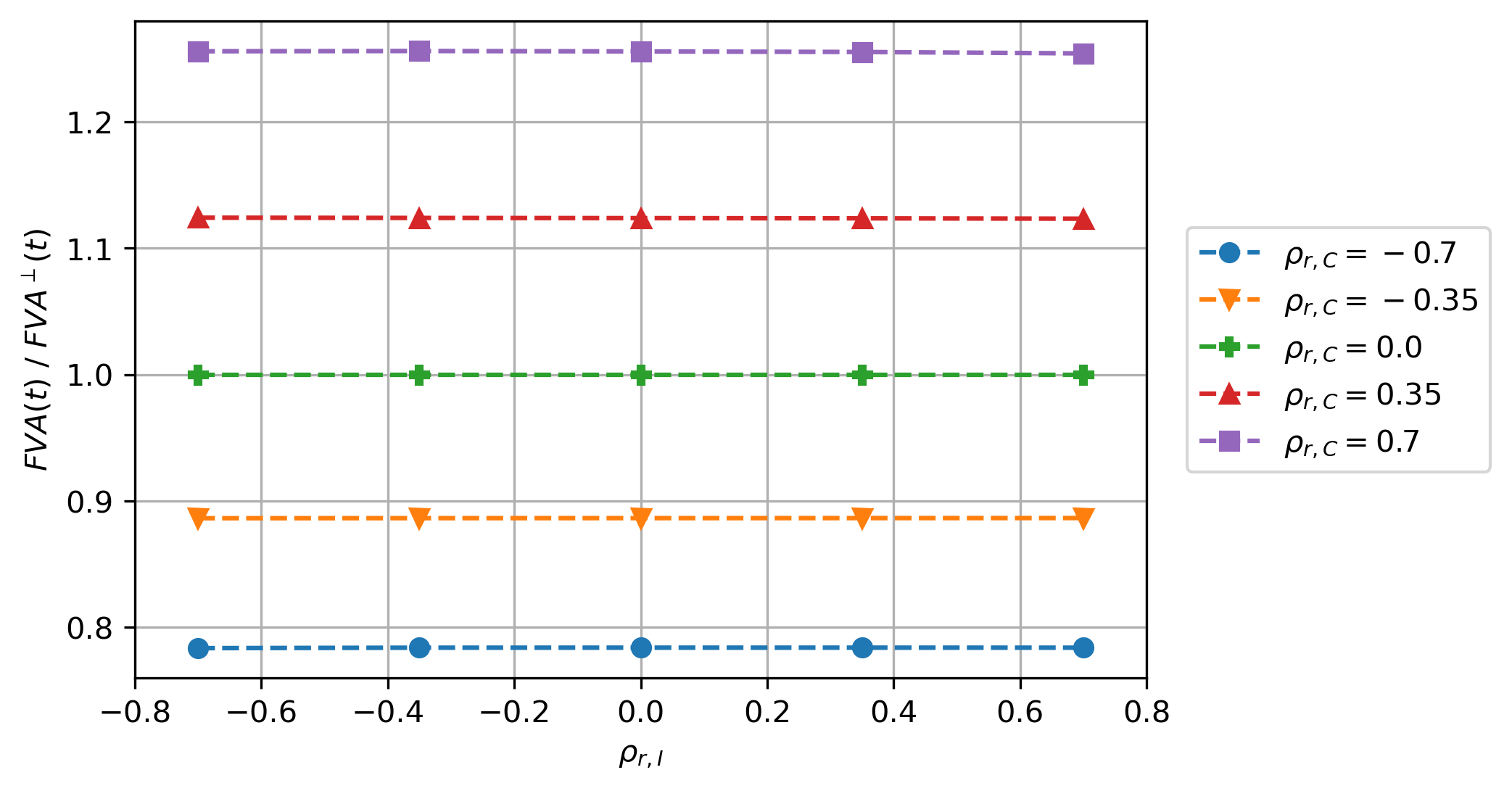}
    \caption{$\default_I$ excl., $\default_C$ incl., $\FVAIndep(t) = 36.11$.}
    \label{fig:corrIRReceiver20ExclIncl}
  \end{subfigure}
  \begin{subfigure}[b]{\resultFigureSize}
    \includegraphics[width=\linewidth]{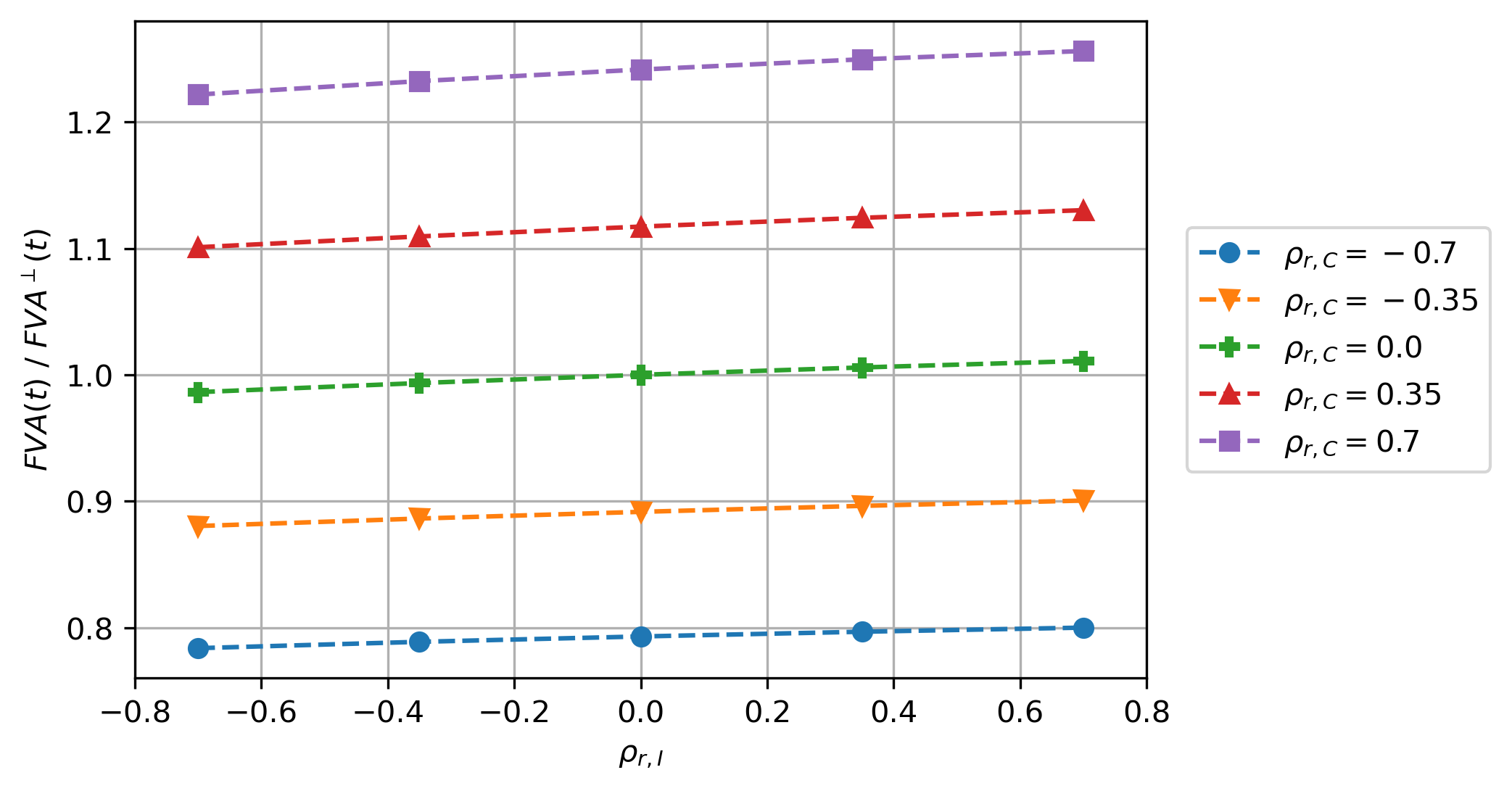}
    \caption{$\default_I$ incl., $\default_C$ incl., $\FVAIndep(t) = 33.72$.}
    \label{fig:corrIRReceiver20InclIncl}
  \end{subfigure}
  \caption{WWR effects for a deterministic spread and an ATM receiver swap.}
  \label{fig:corrIRReceiver20}
\end{figure}

Similarly to the stochastic funding spread results from Section~\ref{sec:wwrRelevanceCreditBasedSpread}, now consider the deterministic funding spread results in Figure~\ref{fig:corrIRReceiver20}.
Now, only RWR effects coming from the credit adjustment factors are expected.
In Figure~\ref{fig:corrIRReceiver20ExclExcl}, when both $\default_I$ and $\default_C$ are excluded, there is neither WWR nor RWR present.

The expected RWR for $I$'s credit adjustment factor is clearly observed in Figure~\ref{fig:corrIRReceiver20InclExcl}, and RWR is roughly linear in $\corr_{\shortRate,I}$.
The RWR is limited with a maximum effect of around $3\%$, explained by the relatively high credit quality of $I$ in this example.

When only including $\default_C$, see Figure~\ref{fig:corrIRReceiver20ExclIncl}, there is again a roughly linear RWR effect, but much more pronounced, resulting from the lower credit quality for $C$.
Unlike the credit-spread case, the various values of $\corr_{\shortRate,I}$ have no impact on the $\FVA$ value.

When combining both credit adjustment factors in Figure~\ref{fig:corrIRReceiver20InclIncl}, as expected, the RWR effect from including $\default_C$ dominates the results.
Though there is only RWR, this is non-negligible.

Looking at the correlation magnitude, the same conclusions can be drawn as for the stochastic spread: when turning off one of the two correlations, the effect is roughly linear; when both correlations are non-zero, the effects are non-trivial.

\subsection{Market conditions} \label{sec:additionalResultsMarketConditions}
Next, the impact of different market conditions and combinations of model parameters is assessed.
We look at the effects of IR level (yield curve), IR volatility, credit level (credit curve), credit volatility and correlation parameters on the WWR for an ATM receiver swap.
The starting point is scenario~\ref{app:scenario2} in~\ref{app:scenarios}, and in scenarios~\ref{app:scenario3}--\ref{app:scenario21} we assess all effects in isolation.
We also look at payer swaps and the three types of moneyness~\footnote{ATM (at-the-money), ITM (in-the-money) and OTM (out-of-the-money).} by calculating the par swap rate and applying up and downward shocks of 50 basis points to the par rate.

The IR parameter impacts are summarized in Table~\ref{tab:wwrEffectsIR}.
When doubling the IR implied volatility (scenario~\ref{app:scenario3}), the $\FVAIndep$ increases with roughly a factor 2 as well.
For an asymmetric funding curve, an increased IR volatility results in a wider exposure profile.
When IR volatilities are correlated with the credit level, this means WWR in the asymmetric funding case.

\begin{table}[h]
  \setlength{\extrarowheight}{1pt}
  \centering
  \footnotesize
  \begin{tabular}{p{2.4cm} l | p{0.65cm} p{0.68cm} p{0.79cm} | p{0.5cm} p{0.68cm} p{0.79cm} | p{0.68cm} p{0.79cm}}
                                                          &                                       & \multicolumn{3}{l|}{Generic} & \multicolumn{3}{p{2.5cm}|}{WWR/RWR for stochastic spread}                 & \multicolumn{2}{p{2.0cm}}{RWR for deterministic spread}             \\
                                                         &                      &             &        &        &        &        &        &        &        \\
   Change in param.                                      & Scenario                & $\FVAIndep$ & I disc & C disc & $\borrowingSpread$ & I disc & C disc & I disc & C disc \\ \hline
   $\vol_{\shortRate} \ \uparrow$                        & \ref{app:scenario3}  & +           & o          & o          & +                           & +               & +               & +               & +               \\
   $a_{\shortRate} \ \uparrow$                           & \ref{app:scenario4}  & -           & o          & o          & -                           & -               & -               & -               & -               \\
   $a_{\shortRate},\vol_{\shortRate} \ \uparrow$         & \ref{app:scenario5}  & +           & o          & o          & +                           & -               & -               & +               & +               \\
   EUR1D curve                                     & \ref{app:scenario6}  & -           & +          & +          & -                           & -               & -               & +               & +               \\
  \end{tabular}
  \caption{WWR/RWR effects under different IR scenarios, for both $\default_I$ and $\default_C$ included in the $\FVA$ definition.
  Signs +/o/- indicates that the quantity has gone up / stayed roughly the same / has gone down.
  }
  \label{tab:wwrEffectsIR}
\end{table}

An increased IR mean reversion, $a_{\shortRate}$, results in a decrease in $\FVAIndep$, and slightly less WWR (scenario~\ref{app:scenario4}).
The effects of changing $\vol_{\shortRate}$ and $a_{\shortRate}$ are multiplicative (scenario~\ref{app:scenario5}).
In this case, overall the volatility effect dominates and $\FVAIndep$ increases.

Changing to the EUR1D curve with negative interest rates (scenario~\ref{app:scenario6}) implies a change in IR level.
In this case, the credit adjustment factor and WWR/RWR effects of including $\default_I$ and/or $\default_C$ are affected.
These results are case-specific, non-trivial and no generic conclusion can be drawn.

The credit parameter impacts are summarized in Table~\ref{tab:wwrEffectsCredit}.
A lower credit level (lower rating credit curve) results in a stronger credit adjustment effect of including $\default_I$ and/or $\default_C$.
This is due to a lower credit quality which translates into lower survival probabilities (scenario~\ref{app:scenario7}).
For a lower credit level of I (scenario~\ref{app:scenario7}), the average funding spread goes up, as this is based on I's credit.
As a result, the $\FVAIndep$ and WWR/RWR effects go up significantly.
$\FVAIndep$ going up in this case can be understood as follows: as credit goes down, the hazard rate goes up, so the funding spread goes up.
For a lower credit level of C (scenarios~\ref{app:scenario8} and~\ref{app:scenario10}), there is a larger credit adjustment effect of including $\default_C$.

\begin{table}[h]
  \setlength{\extrarowheight}{1pt}
  \centering
  \footnotesize
  \begin{tabular}{p{2.4cm} l | p{0.65cm} p{0.68cm} p{0.79cm} | p{0.5cm} p{0.68cm} p{0.79cm} | p{0.68cm} p{0.79cm}}
                                                         &                                       & \multicolumn{3}{l|}{Generic} & \multicolumn{3}{p{2.5cm}|}{WWR/RWR for stochastic spread}                 & \multicolumn{2}{p{2.0cm}}{RWR for deterministic spread}             \\
                                                         &                      &             &        &        &        &        &        &        &        \\
   Change in param.                                      & Scenario                & $\FVAIndep$ & I disc & C disc & $\borrowingSpread$ & I disc & C disc & I disc & C disc \\ \hline
   \multirow{2}{2.5cm}{BBB curve for I and C with $\vol_z \ \downarrow$}      & \ref{app:scenario7}  & +           & +          & +          & +                           & +               & +               & +               & +               \\
                                                         &                      &             &        &        &        &        &        &        &        \\
   \multirow{2}{2.5cm}{BBB curve for C with $\vol_C \ \downarrow$}            & \ref{app:scenario8}  & o           & o          & +          & o                           & o               & +               & o               & +               \\
                                                         &                      &             &        &        &        &        &        &        &        \\
   \multirow{2}{2.5cm}{BBB curve for C with similar $\vol_C$}                 & \ref{app:scenario9}  & o           & o          & o          & o                           & o               & +               & o               & +               \\
                                                         &                      &             &        &        &        &        &        &        &        \\
   \multirow{2}{2.5cm}{B curve for C with similar $\vol_C$}                   & \ref{app:scenario10} & o           & o          & +          & o                           & o               & +               & o               & +               \\
                                                         &                      &             &        &        &        &        &        &        &        \\
   \multirow{2}{2.5cm}{B curve for C with $\vol_C \ \uparrow$}                & \ref{app:scenario11} & o           & o          & +          & o                           & o               & +               & o               & +               \\
                                                         &                      &             &        &        &        &        &        &        &        \\
   $a_I \ \uparrow$                                      & \ref{app:scenario12} & o           & o          & o          & -                           & -               & -               & -               & -               \\
   $\vol_I \ \uparrow$                                   & \ref{app:scenario13} & +           & o          & o          & +                           & +               & o               & +               & o               \\
   $a_I, \vol_I \ \uparrow$                              & \ref{app:scenario14} & +           & o          & o          & +                           & +               & o               & +               & o               \\
   $a_C \ \uparrow$                                      & \ref{app:scenario15} & o           & o          & o          & o                           & o               & -               & o               & -               \\
   $\vol_C \ \uparrow$                                   & \ref{app:scenario16} & o           & o          & o          & o                           & o               & +               & o               & +               \\
   $a_C, \vol_C \ \uparrow$                              & \ref{app:scenario17} & o           & o          & o          & o                           & o               & +               & o               & +               \\
  \end{tabular}
  \caption{WWR/RWR effects under different credit scenarios, for both $\default_I$ and $\default_C$ included in the $\FVA$ definition.
  Signs +/o/- indicates that the quantity has gone up / stayed roughly the same / has gone down.
  }
  \label{tab:wwrEffectsCredit}
\end{table}

Increasing the mean reversion $a_I$ (scenario~\ref{app:scenario12}) results in a smaller $\FVAIndep$.
This comes from the choice of funding spread, which is based on $\intensity_I$.
There is smaller RWR from the inclusion of $\default_I$ and/or $\default_C$.
Increasing the credit volatility $\vol_I$ (scenario~\ref{app:scenario13}) results in an increased $\FVAIndep$ and increased WWR effect.
The effects of changing $\vol_I$ and $a_I$ are multiplicative (scenario~\ref{app:scenario14}).
In this case, overall the volatility effect dominates and $\FVAIndep$ increases.

When increasing the mean reversion $a_C$ (scenario~\ref{app:scenario15}), only the RWR effect from including $\default_C$ is affected: the amount of RWR goes down.
Increased credit volatility $\vol_C$ for the same credit curve (scenarios~\ref{app:scenario9}, \ref{app:scenario11} and~\ref{app:scenario16}) does not seem to change the credit adjustment effect of including $\default_C$.
Hence, the credit adjustment effects are purely driven by the credit curve itself.
Similarly to institution $I$, the effects of the changes in $a_C$ and $\vol_C$ are multiplicative (scenario~\ref{app:scenario17}).
In this example, the effect of the change in $\vol_C$ dominates.

The correlation parameter impacts are summarized in Table~\ref{tab:wwrEffectsCorrelation}.
Changes in correlation magnitude or sign do not affect the level of $\FVAIndep$ or the credit adjustment effects of including $\default_I$ and/or $\default_C$.

\begin{table}[h]
  \setlength{\extrarowheight}{1pt}
  \centering
  \footnotesize
  \begin{tabular}{p{2.4cm} l | p{0.65cm} p{0.68cm} p{0.79cm} | p{0.5cm} p{0.68cm} p{0.79cm} | p{0.68cm} p{0.79cm}}
                                                         &                                       & \multicolumn{3}{l|}{Generic} & \multicolumn{3}{p{2.5cm}|}{WWR/RWR for stochastic spread}                 & \multicolumn{2}{p{2.0cm}}{RWR for deterministic spread}             \\
                                                         &                      &             &        &        &        &        &        &        &        \\
   Change in param.                                      & Scenario                & $\FVAIndep$ & I disc & C disc & $\borrowingSpread$ & I disc & C disc & I disc & C disc \\ \hline
   $\corr_{\shortRate,I} \times 2$                       & \ref{app:scenario18} & o           & o          & o          & +                           & o               & -               & +               & o               \\
   $\corr_{\shortRate,C} \times 2$                       & \ref{app:scenario19} & o           & o          & o          & o                           & o               & +               & o               & +               \\
   $\corr_{\shortRate,I},\corr_{\shortRate,C} \times 2$  & \ref{app:scenario20} & o           & o          & o          & +                           & o               & o               & +               & +               \\
   $\corr_{\shortRate,I},\corr_{\shortRate,C} \times -1$ & \ref{app:scenario21} & o           & o          & o          & - (-)                       & - (-)           & - (-)           & - (-)           & - (-)
  \end{tabular}
  \caption{WWR/RWR effects under different correlation scenarios, for both $\default_I$ and $\default_C$ included in the $\FVA$ definition.
  Signs +/o/- indicates that the quantity has gone up / stayed roughly the same / has gone down.
  `(-)' refers to a change in sign.}
  \label{tab:wwrEffectsCorrelation}
\end{table}

Doubling the magnitude of $\corr_{\shortRate,I}$ (scenario~\ref{app:scenario18}) results in around twice as much WWR from the funding spread.
When excluding $\default_I$ and $\default_C$, the WWR effect from the funding spread is approximately linear in correlation magnitude.
On the other hand, doubling the magnitude of $\corr_{\shortRate,C}$ (scenario~\ref{app:scenario19}) results in a larger RWR effect from including $\default_C$.
When doubling the magnitude of both $\corr_{\shortRate,I}$ and $\corr_{\shortRate,C}$ (scenario~\ref{app:scenario20}), there is approximately twice as much WWR from the funding spread.
Relatively speaking, the RWR effects are the same as before the correlation magnitude increases.
If either $\corr_{\shortRate,I}=0$ or $\corr_{\shortRate,C}=0$, an approximate linear effect in correlation can be observed.
Furthermore, if $\corr_{\shortRate,I},\corr_{\shortRate,C} \neq 0$, the correlation effects are non-trivial due to the mixing of effects.

Similar to the observations from Section~\ref{sec:wwrRelevance}, changing the signs of $\corr_{\shortRate,I}$ and $\corr_{\shortRate,C}$ to positive correlations (scenario~\ref{app:scenario21}), results in a change in WWR sign, i.e., WWR changes into RWR and vice versa.
All the effects are relatively of a lower magnitude than for negative correlations.
Furthermore, the WWR/RWR results are non-symmetric in correlation sign.

For the various moneyness types compared to the ATM case we find that ITM (OTM) makes $\FVAIndep$ go up (down) and the percentage of WWR/RWR go down (up).
Also, ITM (OTM) results in a lower (higher) credit adjustment effect compared to the ATM case.
Furthermore, ITM (OTM) causes the WWR from the funding spread to go down (up) in relative sense w.r.t. the ATM case.

Changing to a payer swap causes RWR to change into WWR and vice versa.
Furthermore, all WWR/RWR effects are lower in magnitude for payer swaps compared to receiver swaps in this particular example.
Hence, the WWR/RWR effects are non-symmetric in swap-type.

For payer swaps, the WWR from the funding spread is much closer to being symmetric in correlation sign than for receiver swaps.
Hence, the non-linearities have less impact for payer swaps.
All other correlation and moneyness effects described above are invariant in swap type.

\section{Conclusion}  \label{sec:conclusion}
We have shown that WWR effects are non-negligible and play an important role in $\FVA$ modelling, especially from a risk-management perspective when dealing with cross-gamma risks.
We wanted to understand $\FVA$ WWR and how it is affected by different modelling choices.
The model reproduced the WWR effects observed in the March 2020 market moves.
The modelling choices impact the $\FVA$ levels and the dependency structure significantly.
We split the $\FVA$ equation into two parts, an independent and a WWR part, to examine the WWR effects in isolation.
There is a substantial credit adjustment effect from adding the possible default times in the $\FVA$ model, where we have seen examples of a $70\%$ reduction in $\FVA$.
For lower credit quality, this effect increases.
In practice, the choice of including the possible default times depends on an institution's preferences.
Yet, portfolio $\FVA$ computation and trade level attribution should always be possible under these modelling choices.
This can be challenging, and requires further research.

The stochastic funding spread generates WWR, while the credit adjustment effects translate into RWR (for a receiver swap and $\corr_{\shortRate,I},\corr_{\shortRate,C} < 0$).
Depending on correlations, credit parameters, IR parameters and product type, the net result is WWR or RWR.
Changing the correlation sign causes WWR to change into RWR and vice versa.
For a deterministic funding spread, there is only RWR coming from the credit adjustment factors.
The credit adjustment effect is similar for all funding spreads.
That the model can only generate RWR in this case means that the WWR as observed in the March 2020 market turmoil is not captured under this specific set of assumptions.
Rather than exhibiting WWR, the effect is completely opposite.

Furthermore, we analyzed the different impacts the market and model parameters can have on the various components of the modelling.
For IR and credit we see that the effects of $\vol_z$ and $a_z$ are multiplicative, but in these examples the volatility effects dominate.
Furthermore, the credit adjustment effects are unaffected by $\vol_z$ and $a_z$, but driven by the IR and credit levels.
In particular, a lower credit level results in a stronger credit adjustment effect.

While much attention is given to the inclusion of the default times in the $\FVA$ model, the correlation parameters remain a key component of the model.
Correlations do not affect $\FVAIndep$ and the credit adjustment effects, but only the WWR/RWR results.
For both funding spread assumptions, the correlation effects in isolation are roughly linear, but when combined, non-linearities are introduced.
In addition, the WWR/RWR is non-symmetric in correlation sign.

This absence of symmetry also holds for different swap types.
Furthermore, like for a change in correlation sign, a change in swap type results WWR to change sign.
The swap's moneyness affects everything: $\FVAIndep$, credit adjustment effects and WWR/RWR effects.
This is because moneyness affects the exposures, which is naturally present in all the aforementioned items.

\begin{figure}[h]
  \centering
  \includegraphics[scale=0.5]{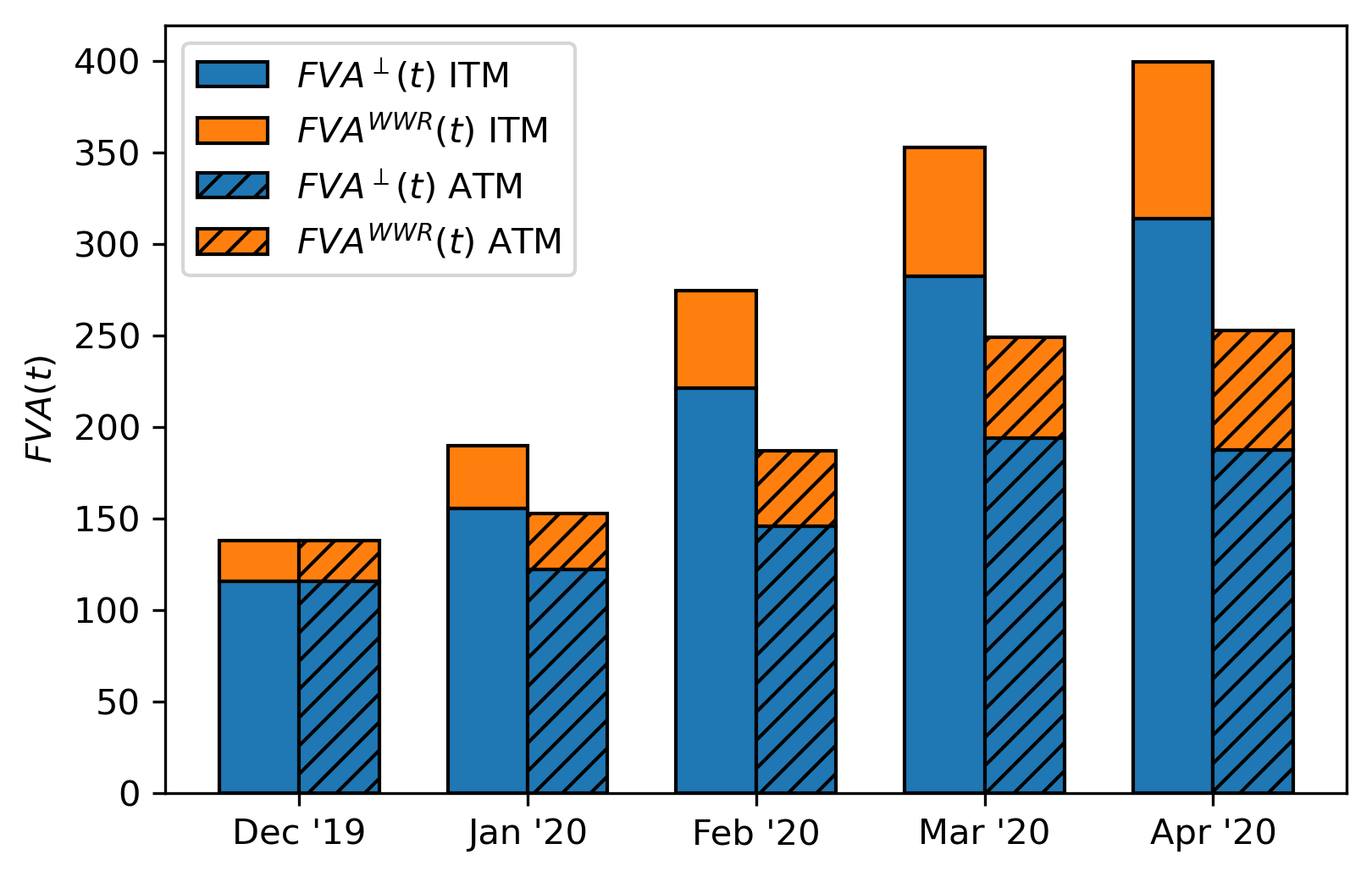}
  \caption{The same situation as in Figure~\ref{fig:FVAThroughTimeITM}, but compared with a similar ATM trade at all dates.
  If at each point in time a portfolio is rebalanced such that it is ATM rather than ITM, the increase through time of overall $\FVA$ is significantly less, but relatively the $\FVAWWR$ becomes more important.}
  \label{fig:FVAThroughTimeComparison}
\end{figure}

The conclusions for the single IR derivative naturally extend to an ITM portfolio of $\FVA$ sensitive trades.
Due to the long-lasting low IR environment, such a portfolio becomes slowly less ITM.
As the new trades continue to stay at-market, the ITM effect slowly fades away, though this might take a long time, especially when there are many long-dated ITM trades in the portfolio.
If interest rates would go up in the future, this would mean a completely opposite situation where the portfolio is OTM.
WWR effects will always strongly depend on the portfolio composition, see Figure~\ref{fig:FVAThroughTimeComparison}.
Actively adding products so that the portfolio is less ITM results in lower $\FVA$ variability, but this may be costly and not always feasible.

We have focused on $\FVA$ WWR in a qualitative sense.
Yet, it is unclear how to compute this quantity efficiently, as a Monte Carlo approach is too expensive in practice.
For each counterparty, an additional credit process needs to be simulated.
As $I$ likely has many counterparties, simulating even more risk-factors is undesired.
Hence, the industry needs a new efficient method to compute $\FVA$ WWR.
Our detailed quantitative approach is part of a forthcoming paper.

\section*{Acknowledgements} This work has been financially supported by Rabobank.
The authors are grateful for valuable discussions with Jasper van der Kroft regarding the contents of this paper.
Furthermore, the authors would like to thank the two anonymous referees for their helpful comments and feedback, which contributed to an improvement of the manuscript.

{\footnotesize
\bibliographystyle{abbrv}
\bibliography{bib/MacroStrings,bib/Articles,bib/Books,bib/Misc,bib/Regulation} 

\begin{thebibliography}{10}

\bibitem{AlbaneseAndersenIabichino201502}
C.~Albanese, L.~Andersen, and S.~Iabichino.
\newblock {FVA accounting, risk management and collateral trading}.
\newblock {\em Risk}, February 2015.

\bibitem{Risk20200416}
L.~Becker.
\newblock {FVA losses back in spotlight after coronavirus stress}.
\newblock Risk.net,
  \url{https://www.risk.net/derivatives/7526696/fva-losses-back-in-spotlight-after-coronavirus-stress},
  April 2020.

\bibitem{Risk20200903}
L.~Becker.
\newblock {XVA traders have no time to rest on laurels}.
\newblock Risk.net,
  \url{https://www.risk.net/our-take/7674091/xva-traders-have-no-time-to-rest-on-laurels},
  September 2020.

\bibitem{BieleckiRutkowski2002}
T.~Bielecki and M.~Rutkowski.
\newblock {\em {Credit Risk: Modeling, Valuation and Hedging}}.
\newblock Springer--Verlag, first edition, 2002.
\newblock {ISBN 978-3-540-67593-8}.

\bibitem{BrigoCapponiPallaviciniPapatheodorou201101}
D.~Brigo, A.~Capponi, A.~Pallavicini, and V.~Papatheodorou.
\newblock {Collateral Margining in Arbitrage-Free Counterparty Valuation
  Adjustment including Re-Hypothecation and Netting}.
\newblock {\em arXiv Electronic Journal}, January 2011.

\bibitem{BrigoFrancischelloPallavicini201904}
D.~Brigo, M.~Francischello, and A.~Pallavicini.
\newblock {Nonlinear valuation under credit, funding, and margins: Existence,
  uniqueness, invariance, and disentanglement}.
\newblock {\em European Journal of Operational Research}, 274(2):788--805,
  April 2019.

\bibitem{BrigoLiuPallaviciniSloth201612}
D.~Brigo, Q.~Liu, A.~Pallavicini, and D.~Sloth.
\newblock {Nonlinearity Valuation Adjustment - Nonlinear Valuation Under
  Collateralization, Credit Risk and Funding Costs}.
\newblock {\em Innovations in Derivatives Markets}, 165:3--35, December 2016.

\bibitem{BrigoMercurio2006}
D.~Brigo and F.~Mercurio.
\newblock {\em {Interest Rate Models - Theory and Practice With Smile,
  Inflation and Credit}}.
\newblock Springer--Verlag, second edition, 2006.
\newblock {ISBN 978-3-540-22149-4}.

\bibitem{BrigoPallavicini201405}
D.~Brigo and A.~Pallavicini.
\newblock {Nonlinear consistent valuation of CCP cleared or CSA bilateral
  trades with initial margins under credit, funding and wrong-way risks}.
\newblock {\em Journal of Financial Engineering}, 1(1):1--60, May 2014.

\bibitem{BrigoPallaviciniPapatheodorou201107}
D.~Brigo, A.~Pallavicini, and V.~Papatheodorou.
\newblock {Arbitrage-free valuation of bilateral counterparty risk for
  interest-rate products: impact of volatilities and correlations}.
\newblock {\em Mathematical Finance}, 14(6):773--802, July 2011.

\bibitem{BurgardKjaer201111}
C.~Burgard and M.~Kjaer.
\newblock {In the balance}.
\newblock {\em Risk}, 11:72--75, November 2011.

\bibitem{BurgardKjaer201210a}
C.~Burgard and M.~Kjaer.
\newblock {The FVA debate: in theory and practice}.
\newblock {\em SSRN Electronic Journal}, October 2012.

\bibitem{Castagna201208}
A.~Castagna.
\newblock {Yes, FVA is a Cost for Derivatives Desks}.
\newblock {\em SSRN Electronic Journal}, August 2012.

\bibitem{ISDA200109}
D'Hulster.
\newblock {Re: Calculation of regulatory capital for counterparty risk.}
\newblock September 2001.
\newblock ISDA.

\bibitem{Green201511}
A.~Green.
\newblock {\em {XVA: Credit, Funding and Capital Valuation Adjustments}}.
\newblock Wiley Finance, first edition, November 2015.
\newblock {ISBN 978-1-118-55678-8}.

\bibitem{Gregory202007}
J.~Gregory.
\newblock {\em {The xVA Challenge - Counterparty Risk, Funding, Collateral,
  Capital and Initial Margin}}.
\newblock Wiley Finance, fourth edition, July 2020.
\newblock {ISBN 978-1-119-50897-7}.

\bibitem{HullWhite201208}
J.~Hull and A.~White.
\newblock {The FVA debate}.
\newblock {\em Risk}, 25:83--85, August 2012.

\bibitem{HullWhite201210}
J.~Hull and A.~White.
\newblock {The FVA debate continued}.
\newblock {\em Risk}, 10:52, October 2012.

\bibitem{HullWhite201409}
J.~Hull and A.~White.
\newblock {Collateral and credit issues in derivatives pricing}.
\newblock {\em The Journal of Credit Risk}, 10(3):3--28, September 2014.

\bibitem{HullWhite201405}
J.~Hull and A.~White.
\newblock {Valuing Derivatives: Funding Value Adjustments and Fair Value}.
\newblock {\em Financial Analysts Journal}, 70(3):46--56, May 2014.

\bibitem{HullWhite201605}
J.~Hull and A.~White.
\newblock {XVAs: A Gap Between Theory and Practice}.
\newblock {\em Risk}, pages 50--52, May 2016.

\bibitem{KenyonBerrahouiPoncet202003}
C.~Kenyon, M.~Berrahoui, and B.~Poncet.
\newblock {Model independent WWR for regulatory CVA and for accounting CVA and
  FVA}.
\newblock {\em arXiv Electronic Journal}, March 2020.

\bibitem{LaughtonVaisbrot201209}
S.~Laughton and A.~Vaisbrot.
\newblock {In defence of FVA - a response to Hull and White}.
\newblock {\em Risk}, 25(9), September 2012.

\bibitem{LongstaffSchwartz200101}
F.~Longstaff and E.~Schwartz.
\newblock {Valuing American Options by Simulation: a Simple Least-quares
  Approach}.
\newblock {\em Review of Financial Studies}, 14(1):113--147, January 2001.

\bibitem{GarciaMunoz201312}
L.~G. Munoz.
\newblock {CVA, FVA (and DVA?) with stochastic spreads. A feasible replication
  approach under realistic assumptions.}
\newblock December 2013.

\bibitem{OosterleeGrzelak201911}
C.~Oosterlee and L.~Grzelak.
\newblock {\em {Mathematical Modeling and Computation in Finance}}.
\newblock World Scientific, first edition, November 2019.
\newblock {ISBN 978-1-78634-794-7}.

\bibitem{PallaviciniPeriniBrigo201112}
A.~Pallavicini, D.~Perini, and D.~Brigo.
\newblock {Funding Valuation Adjustment: a consistent framework including CVA,
  DVA, collateral, netting rules and re-hypothecation}.
\newblock {\em arXiv Electronic Journal}, December 2011.

\bibitem{PallaviciniPeriniBrigo201212}
A.~Pallavicini, D.~Perini, and D.~Brigo.
\newblock {Funding, Collateral and Hedging: uncovering the mechanics and
  subtleties of funding valuation adjustments}.
\newblock {\em arXiv Electronic Journal}, December 2012.

\bibitem{Risk20200824}
N.~Rega-Jones.
\newblock {CVA desks arm themselves for the next crisis}.
\newblock Risk.net,
  \url{https://www.risk.net/derivatives/7669811/cva-desks-arm-themselves-for-the-next-crisis},
  August 2020.

\bibitem{Risk20201105}
N.~Rega-Jones.
\newblock {Funding pain prompts calls to rehome FVA}.
\newblock Risk.net,
  \url{https://www.risk.net/derivatives/7706196/funding-pain-prompts-calls-to-rehome-fva},
  November 2020.

\bibitem{Risk20210121}
R.~Tunstead.
\newblock {The slow corporate embrace of CSAs}.
\newblock Risk.net,
  \url{https://www.risk.net/derivatives/7735746/the-slow-corporate-embrace-of-csas},
  January 2021.

\bibitem{Turlakov201303}
M.~Turlakov.
\newblock {Wrong-way risk, credit and funding}.
\newblock {\em Risk}, 26(3):69--71, March 2013.

\bibitem{Valsecchi202104}
N.~Valsecchi.
\newblock {Funding Value Adjustment and Wrong-Way Risk: the Interest Rate Swap
  case}.
\newblock Master's thesis, Department of Mathematics, Politecnico di Milano,
  April 2021.

\bibitem{ZwaardGrzelakOosterlee202102}
T.~van~der Zwaard, L.~Grzelak, and C.~Oosterlee.
\newblock {A computational approach to hedging Credit Valuation Adjustment in a
  jump-diffusion setting}.
\newblock {\em Applied Mathematics and Computation}, 391, February 2021.

\bibitem{Risk20210309}
L.~Woodall.
\newblock {How XVAs hit top US banks' trading revenues in 2020}.
\newblock Risk.net,
  \url{https://www.risk.net/risk-quantum/7805311/how-xvas-hit-top-us-banks-trading-revenues-in-2020},
  March 2021.

\bibitem{Risk20200611}
L.~Woodall and A.~Bhollah.
\newblock {How XVAs shaped top US dealers' trading revenues in Q1}.
\newblock Risk.net,
  \url{https://www.risk.net/risk-quantum/7560481/how-xvas-shaped-top-us-dealers-trading-revenues-in-q1},
  June 2020.

\end{thebibliography}
}

\setlength\parindent{0pt}
\appendix

\section{The FCA debate} \label{app:fcaDebate}
There has been a debate in literature on the legitimacy of incorporating $\FCA$ in pricing, initiated by Hull and White~\cite{HullWhite201208}.
They argue that $\FCA$ (they call this $\FVA$) should not be considered when valuing a derivative nor when determining the prices charged by the dealer.
According to the authors, the risk-free rate should be used for discounting when valuing derivatives as this is required by the risk-neutral valuation paradigm.
They counter the argument that funding hedging positions would be a natural cause of including $\FCA$, postulating that hedges are zero NPV instruments, and that hence the hedging decision should not impact valuation.
Furthermore, they argue that $\FCA$ is in conflict with the corporate finance theory principle that pricing and funding should be kept segregated: discounting when valuing a project should reflect the risk of the project rather than the risk of the firm.
In conclusion, they state that $\FCA$ is an adjustment that moves away from computing the economic value, and should thus not be included in pricing.

In their original argument, Hull and White~\cite{HullWhite201208} claim that Burgard and Kjaer~\cite{BurgardKjaer201111} make the same point of not including $\FCA$ in valuing or pricing derivatives, but with a different set of arguments.
Burgard and Kjaer state that funding costs are indeed present in practice, as the funding party needs to be compensated for the possibility that the issuer might default on the received funding, and include them in their pricing equations through an $\FCA$ term.
Then, using a simple balance sheet and funding model, they show that when adding the derivative asset and the funding liability to the balance sheet, funding costs can be mitigated and that the $\FCA$ term is reduced to 0 as a result.
As a result, prices are now symmetric between the issuer and the counterparty.
However, from a practical point of view, they argue the operational challenges associated to this theoretical argument, as it requires the ability to completely hedge all risks related to the issuer and counterparty default.
This is only possible when the issuer is able to buy and sell arbitrary quantities of their own bonds, such that the spread earned compensates for the funding costs.
Furthermore, they state some other special cases where the funding adjustment drops out from the valuation.

In a clarifying note~\cite{BurgardKjaer201210a}, Burgard and Kjaer shed some further light on their original arguments and relate them to those of Hull and White~\cite{HullWhite201208}.
They see the $\FCA$ as the expected value of the windfall to the issuer's bondholders at the issuer's default, when the derivative is ITM for the issuer.
The authors argue in which three cases the $\FCA$ term should not be included:
\begin{enumerate}
  \item When the issuer is able to buy and sell arbitrary quantities of their own bonds, such that the spread earned compensates for the funding costs.
  \item When the ITM derivative can be used as collateral to obtain cheaper funding than the otherwise unsecured funding.
  \item When adding the derivative asset and the funding liability to the balance sheet, the issuer's recovery changes and thus the funding rate, resulting in a marginal funding rate equal to the risk-free rate.
\end{enumerate}
All of these cases are only valid from a theoretical perspective though according to the authors, in a practical setting they all have their shortcomings/issues.

The discussion of $\FCA$ evolved by responses from others~\cite{Castagna201208,LaughtonVaisbrot201209}, amongst which practitioners.
They structurally counter the arguments by Hull and White, claiming that the assumptions of the BSM economy no longer hold as the market is incomplete, and always using the risk-free rate for discounting is inappropriate in that case.
Also, the argument of project-specific discounting rather than firm-based discounting is argued against: adding a risk-free hedged portfolio of derivatives does in practice not affect the rate at which the bank can borrow, let alone the fact that existing debt cannot be renegotiated as a result of adding this portfolio.

In turn, Hull and White counter these responses~\cite{HullWhite201210}, casting doubt on the market incompleteness argument, and challenge the justification of $\FCA$ using a hedging argument.
One of their key arguments/assumptions is the idea that when an issuer hedges, it reduces risk and therefore will have a lower funding spread as a result.

Consecutively in \cite{HullWhite201409,HullWhite201605}, Hull and White argue that the return on an investment should be driven by the risk of the investment rather than the average funding cost of the company undertaking the investment.
This relies on the assumption that debtholders of the issuer understand all the risks taken and continuously reflect this in the pricing of debt.
The advocates against including $\FCA$ in pricing are financial economists who work with marginal funding costs, as opposed to financial engineers who are in favour of including $\FCA$ and use average funding costs.
Concerns are expressed about $\FCA$ being an asymmetric adjustment that will jeopardize the ability of two parties to agree on a price for a deal.
Their view is that inclusion of funding costs lacks a theoretical basis, but they recognize the practical situation of institutions including it in pricing.
A possible future scenario according to the authors is $\FCA$ being used for internal decision making when taking on a deal or not.

In~\cite{HullWhite201405}, Hull and White summarize old arguments and give new ones as well.
A key new insight is that funding adjustments are driven by derivatives desk performance measurement: they are required to earn a target funding rate.

\section{FVA derivation} \label{app:fvaDerivation}

The following filtrations are relevant: $\F(t)$ is the `standard' filtration; $\H_I(t) = \sigma\left(\left\{\default_I \leq s \right\}:s \leq t\right)$ is the filtration generated by the default time $\default_I$; $\H_C(t) = \sigma\left(\left\{\default_C \leq s \right\}:s \leq t\right)$ is the filtration generated by the default time $\default_C$; $\G(t) := \F(t) \otimes \H_I(t) \otimes \H_C(t)$ is the enriched filtration containing all available market information.

The independence of defaults can be translated into a useful result for later derivations.
As we model times to default $\default_z$, $z \in \{I,\ C\}$, as the first jumps of a Cox process with hazard rate (intensity) $\intensity_z(t)$, we can write for $s \leq t$~\cite{BrigoMercurio2006}:
\begin{align} \displaystyle
  \Q\left(\left. t < \default_z \right| \F(s)\right)
    &= \condExp{\indicator{t < \default_z} }{s}
    = \condExp{\expPower{-\int_{0}^{t} \intensity_z(v)\dv} }{s}. \label{eq:pd1}
\end{align}
Combining this result with the assumption of independent default times $\default_I$ and $\default_C$, results in the following assumption.
\begin{assumption}[Independent defaults] \label{assump:indepCredit}
We assume that default times $\default_I$ and $\default_C$ are conditionally independent on $\F(t)$, hence $\corr_{I,C} = 0$ and, for $s \leq t$,
\begin{align} \displaystyle
  \Q\left(\left.t < \default_I, t < \default_C \right| \F(s)\right)
    &\stackrel{\ \ \ \ \ }{=} \Q\left(\left.t < \default_I\right| \F(s)\right) \cdot \Q\left(\left. t < \default_C \right| \F(s)\right) \nonumber \\
    &\stackrel{\text{\eqref{eq:pd1}}}{=} \condExp{\expPower{-\int_{0}^{t} \intensity_I(v)\dv}}{s} \cdot \condExp{\expPower{-\int_{0}^{t} \intensity_C(v)\dv}  }{s} \nonumber \\
    &\stackrel{\ \ \ \ \ }{=} \condExp{\expPower{-\int_{0}^{t} \intensity_I(v) + \intensity_C(v)\dv} }{s}.\label{eq:pd2}
\end{align}
\end{assumption}

For $\FVA$, we can derive the following expression starting from its definition~\cite{AlbaneseAndersenIabichino201502}, under the single assumption of conditional independence of defaults:
\begin{align} \displaystyle
  \FVA(t)
    &= \E \left[ \left. \int_t^{T\wedge\default_I\wedge\default_C} \expPower{-\int_{t}^{u} \shortRate(v)\dv} \borrowingSpread(u) \maxOperator{\tradeVal(u)} \du \right| \G(t) \right] \nonumber \\
    &= \E \left[ \left. \int_t^{T} \indicator{t < u < \default_I} \indicator{t < u < \default_C} \expPower{-\int_{t}^{u} \shortRate(v)\dv} \borrowingSpread(u) \maxOperator{\tradeVal(u)} \du \right| \G(t) \right] \nonumber \\
    &\stackrel{\text{Fubini}}{=} \int_t^{T} \E \left[ \left. \indicator{t < u < \default_I} \indicator{t < u < \default_C} \expPower{-\int_{t}^{u} \shortRate(v)\dv} \borrowingSpread(u) \maxOperator{\tradeVal(u)}  \right| \G(t) \right] \du \nonumber \\
    &\stackrel{\text{Lemma~\cite{BieleckiRutkowski2002}}}{=} \indicator{t < \default_I}  \indicator{t < \default_C}  \int_t^{T} \frac{ \condExp{ \indicator{u < \default_I} \indicator{u < \default_C} \expPower{-\int_{t}^{u} \shortRate(v)\dv} \borrowingSpread(u) \maxOperator{\tradeVal(u)}  }{t}}{\Q\left(\left.t < \default_I, t < \default_C \right| \F(t)\right) } \du \nonumber \\
    &\stackrel{\text{\eqref{eq:pd2}}}{=} \frac{\indicator{t < \default_I}  \indicator{t < \default_C}}{\expPower{-\int_{0}^{t} \intensity_I(v) + \intensity_C(v)\dv} }  \int_t^{T} \condExp{\indicator{u < \default_I} \indicator{u < \default_C} \expPower{-\int_{t}^{u} \shortRate(v)\dv} \borrowingSpread(u) \maxOperator{\tradeVal(u)}  }{t} \du, \label{eq:fcaApp1}
\end{align}
where in the last step we also used that $\expPower{-\int_{0}^{t} \intensity_I(v) + \intensity_C(v)\dv}$ is $\F(t)$-measurable.
For convenience, we re-iterate Lemma 5.1.2 from~\cite{BieleckiRutkowski2002} we have used.
See also Section 22.5 from~\cite{BrigoMercurio2006} for further background.
\begin{lem} \label{lem:expGtoF}
Let $\G(t) := \F(t) \otimes \sigma\left(\left\{\default \leq s \right\}:s \leq t\right)$.
When $X$ is an $\F(u)$ integrable random variable then for $t<u$:
\begin{align} \displaystyle
   \E \left[ \left. X \indicator{\tau > u}\right| \G(t) \right]
     &=  \indicator{\tau > t} \frac{\condExp{X \indicator{\tau > u}}{t}}{\condExp{\indicator{\tau > t}}{t}}. \nonumber
\end{align}
\end{lem}

By $\F(t) \subset \F(u)$, we can use the tower property to rewrite the following expectation, using that $\expPower{-\int_{t}^{u} \shortRate(v)\dv} \borrowingSpread(u) \maxOperator{\tradeVal(u)}$ and $\expPower{-\int_{0}^{u} \intensity_I(v) + \intensity_C(v)\dv}$ are $\F(u)$-measurable and using Equation~\eqref{eq:pd2}.
\begin{align} \displaystyle
  &\condExp{\indicator{u < \default_I} \indicator{u < \default_C}\expPower{-\int_{t}^{u} \shortRate(v)\dv} \borrowingSpread(u) \maxOperator{\tradeVal(u)}  }{t} \nonumber \\
    &\stackrel{\text{Tower prop.}}{=}  \condExp{ \condExp{ \indicator{u < \default_I} \indicator{u < \default_C}\expPower{-\int_{t}^{u} \shortRate(v)\dv} \borrowingSpread(u) \maxOperator{\tradeVal(u)} }{u} }{t} \nonumber \\
    &\qquad = \condExp{ \condExp{  \indicator{u < \default_I} \indicator{u < \default_C}}{u}  \expPower{-\int_{t}^{u} \shortRate(v)\dv} \borrowingSpread(u) \maxOperator{\tradeVal(u)}   }{t}  \nonumber \\
    &\qquad = \condExp{  \expPower{-\int_{0}^{u} \intensity_I(v) + \intensity_C(v)\dv}\expPower{-\int_{t}^{u} \shortRate(v)\dv} \borrowingSpread(u) \maxOperator{\tradeVal(u)}   }{t}. \label{eq:fcaAppInterm}
\end{align}

We write $\condExp{\cdot}{t} = \condExpSmall{\cdot}{t}$ for ease of notation.
Furthermore, we assume that no defaults take place before $t$, which in practice will be today.
In other words, we have $\indicator{t < \default_I}  \indicator{t < \default_C} = 1$.
Going back to $\FVA$ from Equation~\eqref{eq:fcaApp1}, we can now write the following result:
\begin{align} \displaystyle
  \FVA(t)
    &\stackrel{\text{\eqref{eq:fcaAppInterm}}}{=} \frac{1}{\expPower{-\int_{0}^{t} \intensity_I(v) + \intensity_C(v)\dv} }  \int_t^{T} \condExpSmall{ \expPower{-\int_{0}^{u} \intensity_I(v) + \intensity_C(v)\dv}\expPower{-\int_{t}^{u} \shortRate(v)\dv} \borrowingSpread(u) \maxOperator{\tradeVal(u)} }{t} \du \nonumber \\
    &= \int_t^{T} \condExpSmall{ \expPower{-\int_{t}^{u} \intensity_I(v) + \intensity_C(v)\dv}\expPower{-\int_{t}^{u} \shortRate(v)\dv} \borrowingSpread(u) \maxOperator{\tradeVal(u)} }{t} \du\nonumber \\
    &\rdef \int_t^{T} \EPEFVA{t}{u} \du,  \label{eq:fcaApp2}
\end{align}
where in the second step we again used that $\expPower{-\int_{0}^{t} \intensity_I(v) + \intensity_C(v)\dv}$ is $\F(t)$-measurable.

\section{Exposure derivation} \label{app:exposureDerivation}

The exposure $\EPEFVAWWR{t}{u}$ in Equation~\eqref{eq:fca1} can be considered in the following generic form:
\begin{align} \displaystyle
  \EPEFVA{t}{u}
    & = \condExpSmall{f(t,u;\intensity_I,\intensity_C) g(t,u;\shortRate) h(t,u;\shortRate,\tradeVal) }{t}. \label{eq:epe1App}
\end{align}
The choices of functions $f(\cdot)$, $g(\cdot)$ and $h(\cdot)$ depend on the borrowing spread assumptions, as well as on the inclusion/exclusion of $\default_I$ and/or $\default_C$ in the $\FVA$ definition~\eqref{eq:fca0}.
In~\cite{KenyonBerrahouiPoncet202003}, a model-independent approach is used for this, where the expectation is decomposed into simpler expectations, standard deviations and correlations.
Unfortunately, this yields rather complicated expressions that are not trivial to compute.
Instead of decomposing into correlations, we decompose into covariances.
Using the covariance definition iteratively, we can write
\begin{align} \displaystyle
  \EPEFVA{t}{u}
    &= \condExpSmall{f}{t} \condExpSmall{g}{t} \condExpSmall{ h}{t} \nonumber \\
    &\quad + \condExpSmall{ h}{t} \condCovSmall{f}{g}{t} + \condExpSmall{\left(h - \condExpSmall{h}{t} \right)\left(f g - \condExpSmall{f}{t} \condExpSmall{g}{t} - \condCovSmall{f}{g}{t}\right)}{t} \nonumber \\
    &\rdef \EPEFVAIndep{t}{u} + \EPEFVAWWR{t}{u}. \label{eq:epe2App}
\end{align}
The initial expectation can be written as the sum of the independent exposure $\EPEFVAIndep{t}{u}$ and a term capturing the cross-dependencies, $\EPEFVAWWR{t}{u}$, which is driven by the correlation assumptions.

For the stochastic funding spread from Equation~\eqref{eq:fundingSpreadCredit}, the functions in Equation~\eqref{eq:epe2App} are: $g(t,u;\shortRate) = 1$, $h(t,u;\shortRate,\tradeVal) = \expPower{-\int_{t}^{u} \shortRate(v)\dv}\maxOperator{\tradeVal(u)}$, such that $\condExpSmall{\expPower{-\int_{t}^{u} \shortRate(v)\dv}\maxOperator{\tradeVal(u)}}{t}$ is simply the discounted positive exposure, readily available from an existing $\xva$ engine.
Finally, define $f(\cdot)$ as:
\begin{align} \displaystyle
    f(t,u;\intensity_I,\intensity_C)
      &= \expPower{-\int_{t}^{u} \intensity_I(v) + \intensity_C(v)\dv} \borrowingSpread(u), \nonumber \\
    \condExpSmall{f(t,u;\intensity_I,\intensity_C)}{t}
      &= \zcb_{I}(t,u) \zcb_{C}(t,u)\mu_{S}(t, u) + \LGD_I \condExpSmall{\expPower{-\int_{t}^{u} \intensity_I(v) + \intensity_C(v)\dv} y_I(t,u)}{t},
      \nonumber
\end{align}
where the survival probabilities $\zcb_{I}(t,u)$ and $\zcb_{C}(t,u)$ are independent due to the assumption of independent defaults of counterparties $I$ and $C$ from Section~\ref{sec:SDE}.

Applying this to the two exposure types in Equation~\eqref{eq:epe2App} yields:
\begin{align} \displaystyle
  \EPEFVAIndep{t}{u}
    &= \zcb_{I}(t,u) \zcb_{C}(t,u)\mu_{S}(t, u)\condExpSmall{\expPower{-\int_{t}^{u} \shortRate(v)\dv}\maxOperator{\tradeVal(u)}}{t}  \nonumber \\
    &\quad + \LGD_I\condExpSmall{\expPower{-\int_{t}^{u} \intensity_I(v) + \intensity_C(v)\dv} y_I(t,u)}{t} \condExpSmall{\expPower{-\int_{t}^{u} \shortRate(v)\dv}\maxOperator{\tradeVal(u)}}{t},\label{eq:epeIndepCreditaApp} \\
  \EPEFVAWWR{t}{u}
    &= \condExpSmall{\left(\expPower{-\int_{t}^{u} \shortRate(v)\dv}\maxOperator{\tradeVal(u)} - \condExpSmall{\expPower{-\int_{t}^{u} \shortRate(v)\dv}\maxOperator{\tradeVal(u)}}{t} \right)\expPower{-\int_{t}^{u} \intensity_I(v) + \intensity_C(v)\dv}\borrowingSpread(u) }{t}. \label{eq:epeWWRCredit1App}
\end{align}

\begin{rem}[Alternative $\FVA$ definition]
The definition of $f(\cdot)$ above is based on the inclusion of both $\default_I$ and $\default_C$ in the $\FVA$ definition.
If both are excluded, the credit adjustment factor term disappears and we have $f(t,u;\intensity_I,\intensity_C) = \borrowingSpread(u)$ with $\condExpSmall{f(t,u;\intensity_I,\intensity_C)}{t} = \mu_{S}(t, u)$, as $\condExpSmall{y_I(t,u)}{t}=0$.
\end{rem}

\begin{rem}[Deterministic funding spread]
For a deterministic funding spread, we take the same definition of $h(\cdot)$, as before.
For $f(\cdot)$ and $g(\cdot)$, we write:
\begin{align} \displaystyle
    \begin{array}{rclrcl}
      f(t,u;\intensity_I,\intensity_C) & = & \expPower{-\int_{t}^{u} \intensity_I(v) + \intensity_C(v)\dv}, & \condExpSmall{f(t,u;\intensity_I,\intensity_C)}{t} & = & \zcb_{I}(t,u) \zcb_{C}(t,u), \\
      g(t,u;\shortRate) & = & \borrowingSpread(u), & \condExpSmall{g(t,u;\shortRate)}{t} & = & \borrowingSpread(u).
    \end{array} \nonumber
\end{align}
\end{rem}

\section{Dependency structure} \label{app:dependencyStructure}

Here, we motivate why we choose to work with correlated Brownian motions with independent defaults of counterparties $I$ and $C$.

\subsection{Linear versus non-linear dependence} \label{app:linVsNonLin}

Correlation is a measure of linear dependence.
Using correlation, we include generic WWR in the modelling, which captures macro-economic effects.
Despite using only linear dependencies, there are non-linearities in WWR/RWR when varying the correlation parameters.
Due to the mixing of correlation effects, the correlation effects on WWR/RWR become non-trivial, see Figure~\ref{fig:corrCreditReceiver20}.

There are various choices for adding a non-linear dependence rather than a linear dependence
For example, a copula with different upper- and lower-tail dependence can be used.
The current approach of correlated Brownian increments with a single correlation parameter results in equal tail dependence.
Hence, it is comparable to using a Gaussian copula.
Alternatively, staying in the setting with correlated dynamics, correlated jumps could be added to the dynamics to add non-linearities.

\subsection{Independence of defaults} \label{app:corrAssump}

In Section~\ref{sec:SDE}, we argue that assuming independent defaults of counterparties $I$ and $C$ (i.e., $\corr_{I,C} = 0$) is justified as this is not the main driver in WWR modelling.
For WWR modelling for IR derivatives, the dependency between the funding spread and the IR exposure will be the main driver.
This basic assumption simplifies the FVA evaluation, since default probabilities can be factorized.
This allows for analytical tractability.
It is a classical assumption in literature when not dealing with credit derivatives.
Here, we use a numerical example to support this claim.

First of all, non-zero $\corr_{I,C}$ will only have an effect if $\default_C$ is included in the $\FVA$ definition.
Otherwise, there is no dependency on $C$'s credit, so there will be no correlation effect.
Furthermore, we only consider $\corr_{I,C} > 0$ as this represents the case of possible default contagion, which would be relevant in, for example, an economic downturn.

\begin{figure}[h]
  \centering
  \includegraphics[scale=0.5]{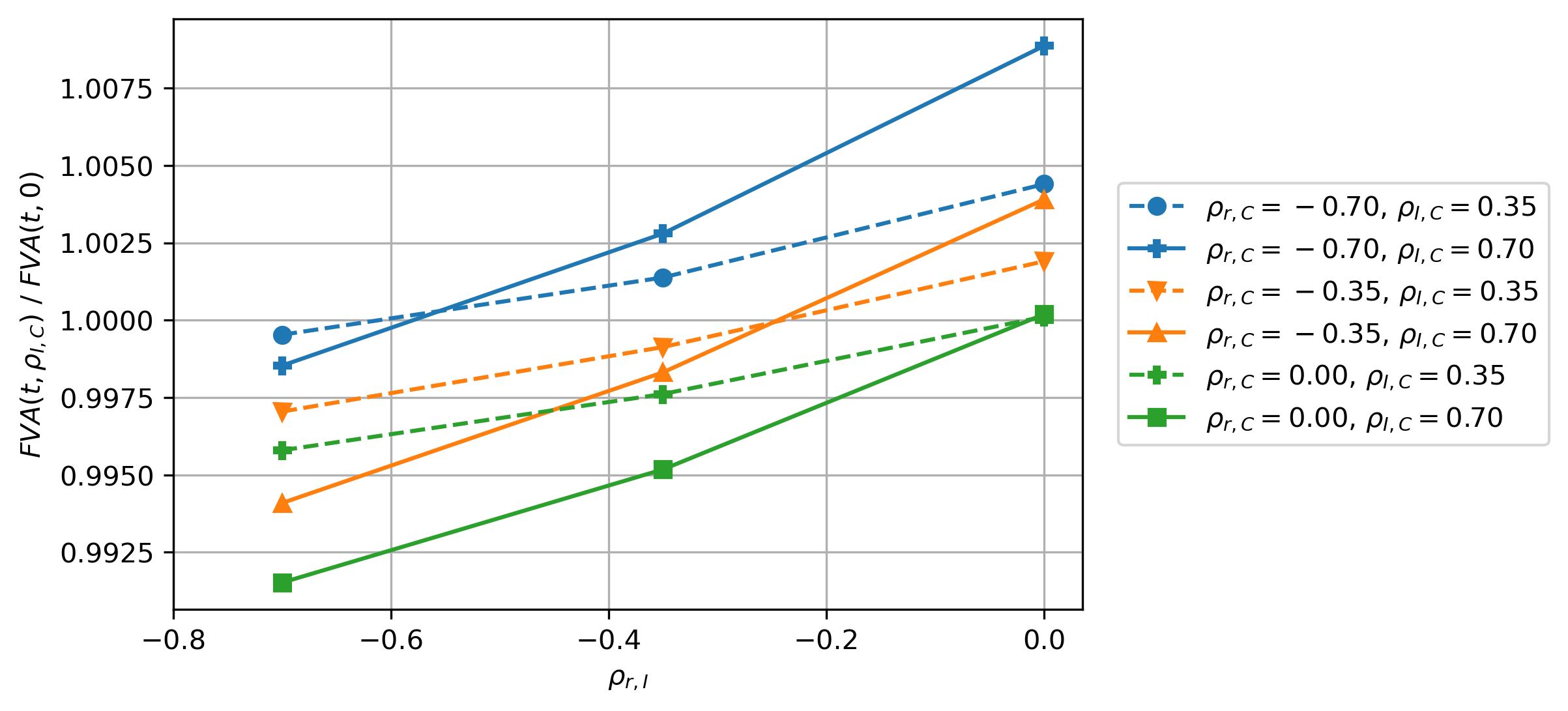}
    \caption{$\frac{\FVA\left(t,\corr_{I,C}\right)}{\FVA(t,0)}$ for a stochastic funding spread, $\default_I$ incl., $\default_C$ incl.}
    \label{fig:corrCreditReceiver20InclInclRhoICRatio}
\end{figure}

In Figure~\ref{fig:corrCreditReceiver20InclInclRhoICRatio}, we plot the $\FVA(t)$ value for a correlation $\corr_{I,C}$, i.e., $\FVA(t,\corr_{I,C})$, divided by the case of zero credit-credit correlation, i.e., $\FVA(t,0)$.
These results correspond to the same example as in Section~\ref{sec:wwrRelevance}, based on the setup described in~\ref{app:marketDataModelParameters}.
From this figure, we conclude that the impact of $\corr_{I,C}$ is insignificant.
Hence, we can safely assume that $\corr_{I,C} = 0$.

In fact, it is known that adding the diffusive correlation (i.e., setting $\corr_{I,C} \neq 0$), only yields a weak dependence between credit spreads~\cite[Section 16.8.3]{Green201511}.
So if credit spread correlation is desired, this needs to be incorporated into the model in a different way.
For example, one could use a copula to correlate the creditworthiness of the two counterparties.
Still, adding this additional dependency to the model is only justified if the outcomes are used for risk-management.

\section{Additional results for modelling assumptions} \label{app:additionalResultsModellingAssumptions}
The first results correspond to the market data and model parameters from scenario~\ref{app:scenario1} in~\ref{app:scenarios}, for an ITM receiver swap and a stochastic funding spread.
We examine the various choices of including $\default_I$ and/or $\default_C$ in the $\FVA$ definition.

\begin{figure}[h]
  \centering
  \begin{subfigure}[b]{\resultFigureSize}
    \includegraphics[width=\linewidth]{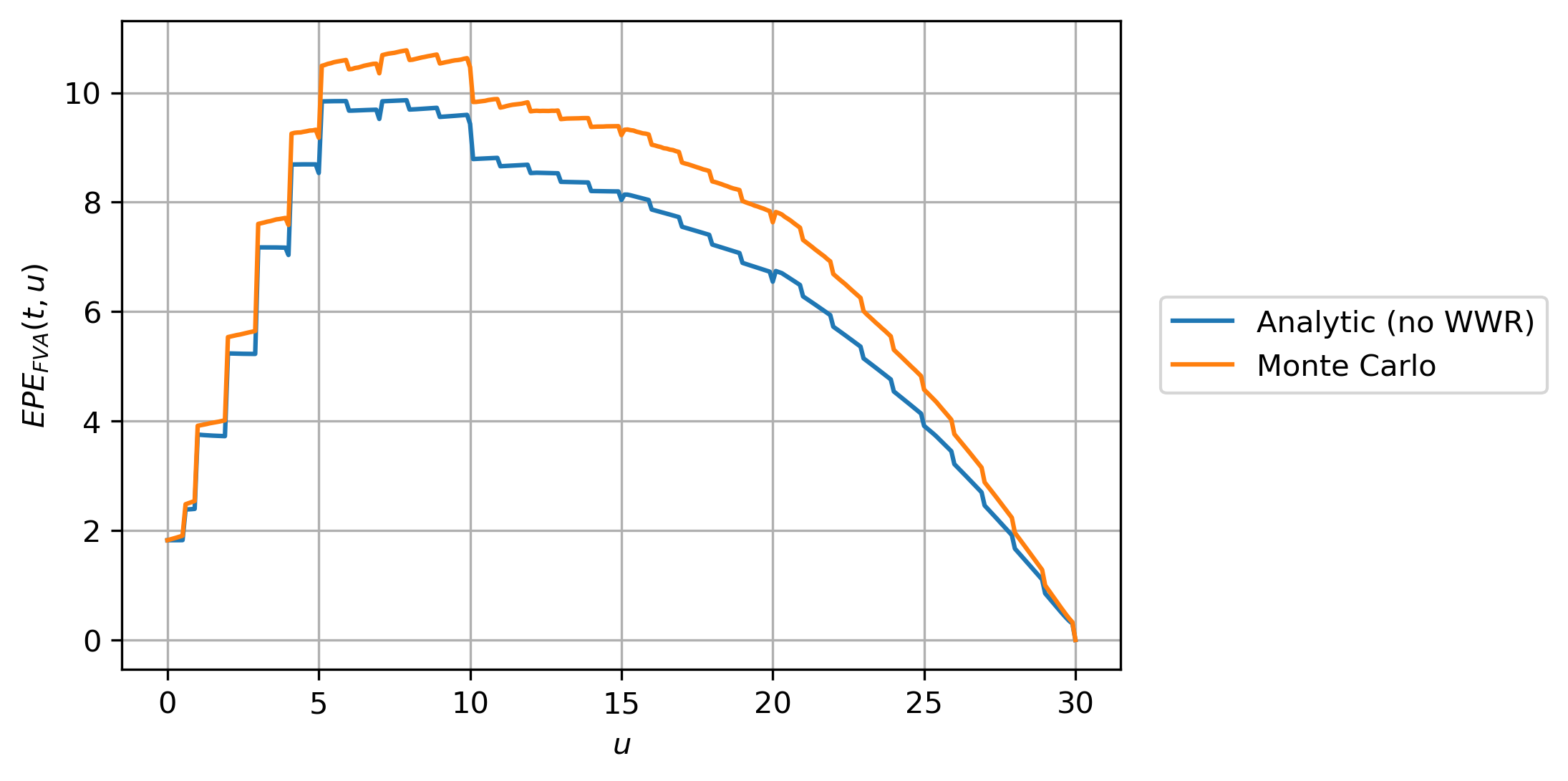}
    \caption{$\EPEFVA{t}{u}$}
    \label{fig:creditITMReceiver31ExclExclExposure}
  \end{subfigure}
  \begin{subfigure}[b]{\resultFigureSize}
    \includegraphics[width=\linewidth]{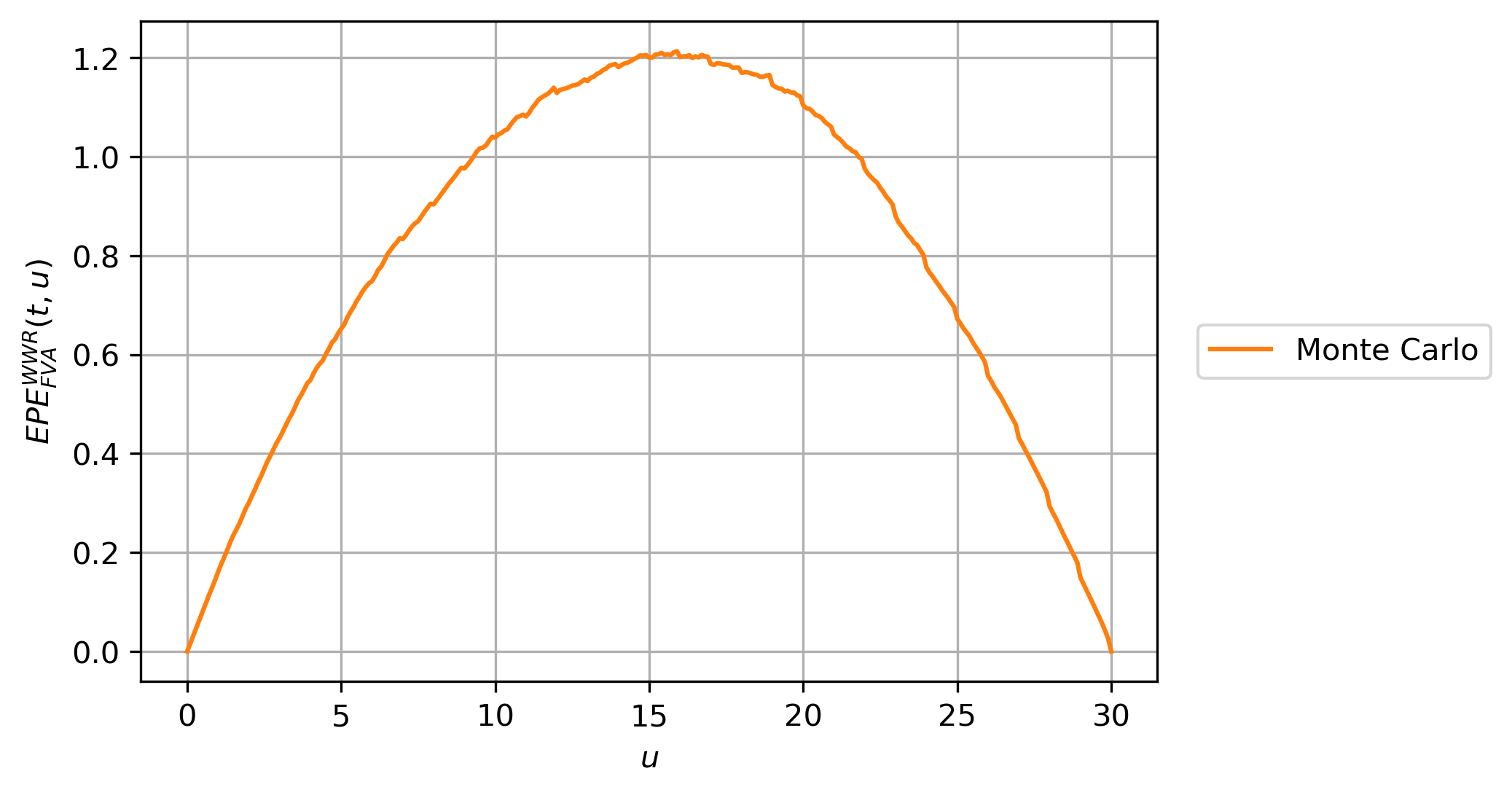}
    \caption{$\EPEFVAWWR{t}{u}$}
    \label{fig:creditITMReceiver31ExclExclWWR}
  \end{subfigure}
  \caption{Stochastic spread, ITM receiver swap, $\default_I$ excl., $\default_C$ excl., $\FVAIndep(t) = 193.3481$.
  The `Analytic' exposure profile is taken from the existing $\xVA$ engine where all quantities are assumed to be independent, and can hence be computed analytically for an IR swap.
  This exposure profile and corresponding $\FVAIndep(t)$ serve as a point of reference to illustrate the amount of WWR that is present after introducing the dependencies.}
  \label{fig:creditITMReceiver31ExclExcl}
\end{figure}

In Figure~\ref{fig:creditITMReceiver31ExclExcl}, exposure plots for $\EPEFVA{t}{u}$ and $\EPEFVAWWR{t}{u}$ are presented for the case that both $\default_I$ and $\default_C$ are excluded from the $\FVA$ definition.
From Figure~\ref{fig:creditITMReceiver31ExclExclExposure}, it is clear that the funding spread $\borrowingSpread$ results in WWR, since the analytic exposure without WWR is below all other exposure profiles.
The same conclusion is drawn from Figure~\ref{fig:creditITMReceiver31ExclExclWWR} where $\EPEFVAWWR{t}{u} > 0$ for all $u$.
Hence, this WWR effect is clearly non-negligible and plays an important role in $\FVA$ modelling.

In Table~\ref{tab:CreditSpreadParamSet31ReceiverITMFvaResults}, the corresponding $\FVA$ and WWR numbers are presented.
These results support the conclusions made from Figure~\ref{fig:creditITMReceiver31ExclExcl}, but now in terms of $\FVA$ numbers rather than exposure profiles.

\begin{table}[h]
  \renewcommand{\arraystretch}{1.1}
  \centering
  \footnotesize
  \begin{tabular}{ll|rrrrr}
     $\default_I$  & $\default_C$   & $\FVAIndep(t)$ & $\FVAWWR(t)$ & $\WWR \%$  \\ \hline
     Excl.& Excl.                   & 193.3481 & 24.0972 & 12.46 \\
     Incl.& Excl.                   & 169.9607 & 18.2658 & 10.75 \\
     Excl.& Incl.                   & 136.5265 & 6.6041  & 4.84 \\
     Incl.& Incl.                   & 122.3386 & 4.7654  & 3.90 \\
  \end{tabular}
  \caption{Stochastic spread, ITM receiver swap.
  `$\WWR \%$' refers to the percentage of $\FVAWWR(t)$ w.r.t. $\FVAIndep(t)$, i.e., $\frac{\FVAWWR(t)}{\FVAIndep(t)} \cdot 100\%$.}
  \label{tab:CreditSpreadParamSet31ReceiverITMFvaResults}
\end{table}

When including $\default_I$ in the $\FVA$ definition, the plots in Figure~\ref{fig:creditITMReceiver31ExclExcl} do not change significantly in shape, but the exposure profile gets scaled down slightly due to the credit adjustment effect.
In Table~\ref{tab:CreditSpreadParamSet31ReceiverITMFvaResults} this credit adjustment is clearly present when looking at the $\FVAIndep$ and $\FVAWWR$ numbers.

In the current situation, the inclusion of $\default_I$ translates into RWR.
This the credit adjustment effect is not the same as RWR, but it is a separate effect.
The credit adjustment affects the $\FVA$ level, while the dependency structure with the existing factors results in the RWR effect, which is seen from the lower percentage of WWR in Table~\ref{tab:CreditSpreadParamSet31ReceiverITMFvaResults}.

\begin{figure}[h]
  \centering
  \begin{subfigure}[b]{\resultFigureSize}
    \includegraphics[width=\linewidth]{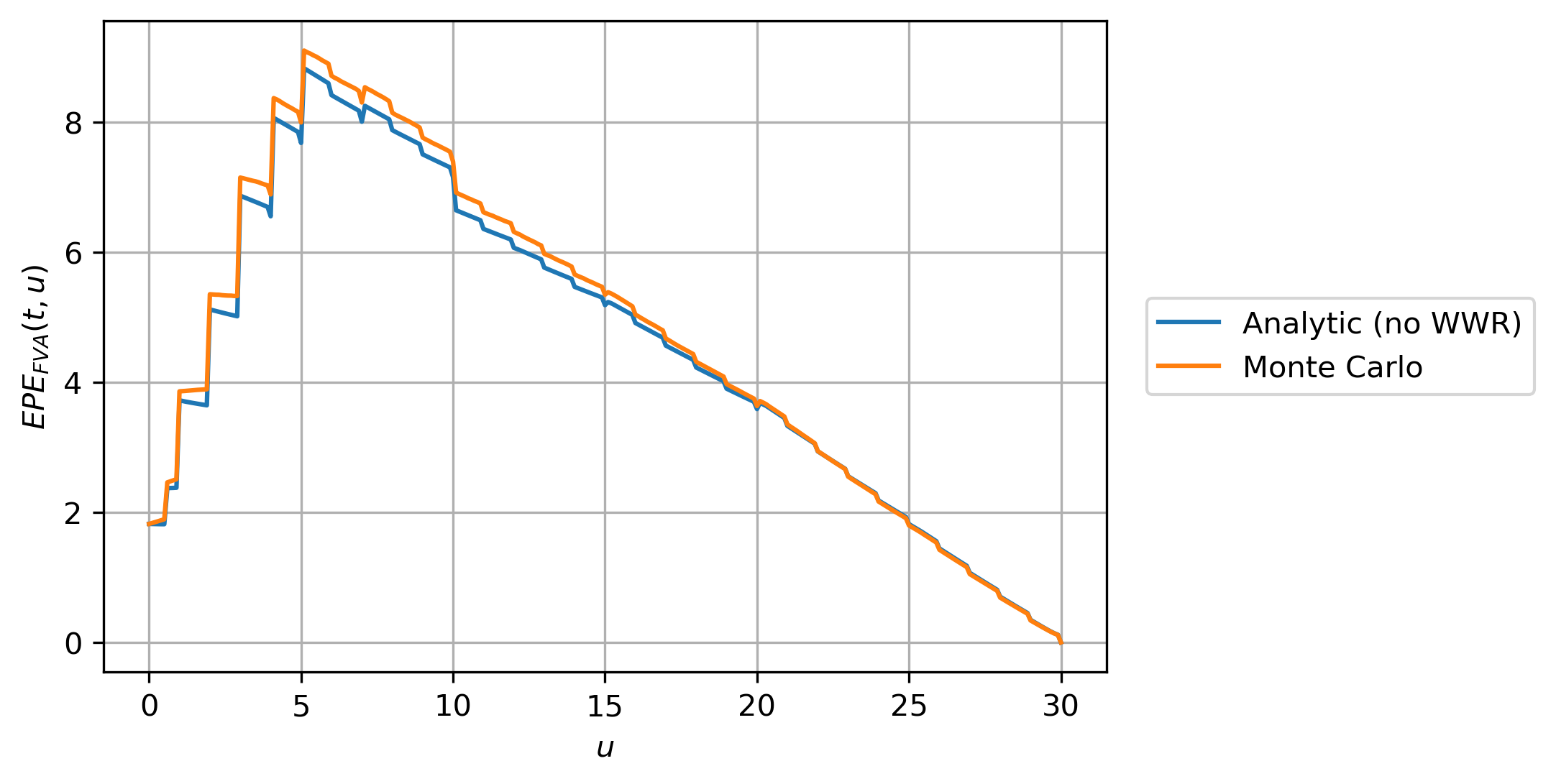}
    \caption{$\EPEFVA{t}{u}$}
    \label{fig:creditITMReceiver31ExclInclExposure}
  \end{subfigure}
  \begin{subfigure}[b]{\resultFigureSize}
    \includegraphics[width=\linewidth]{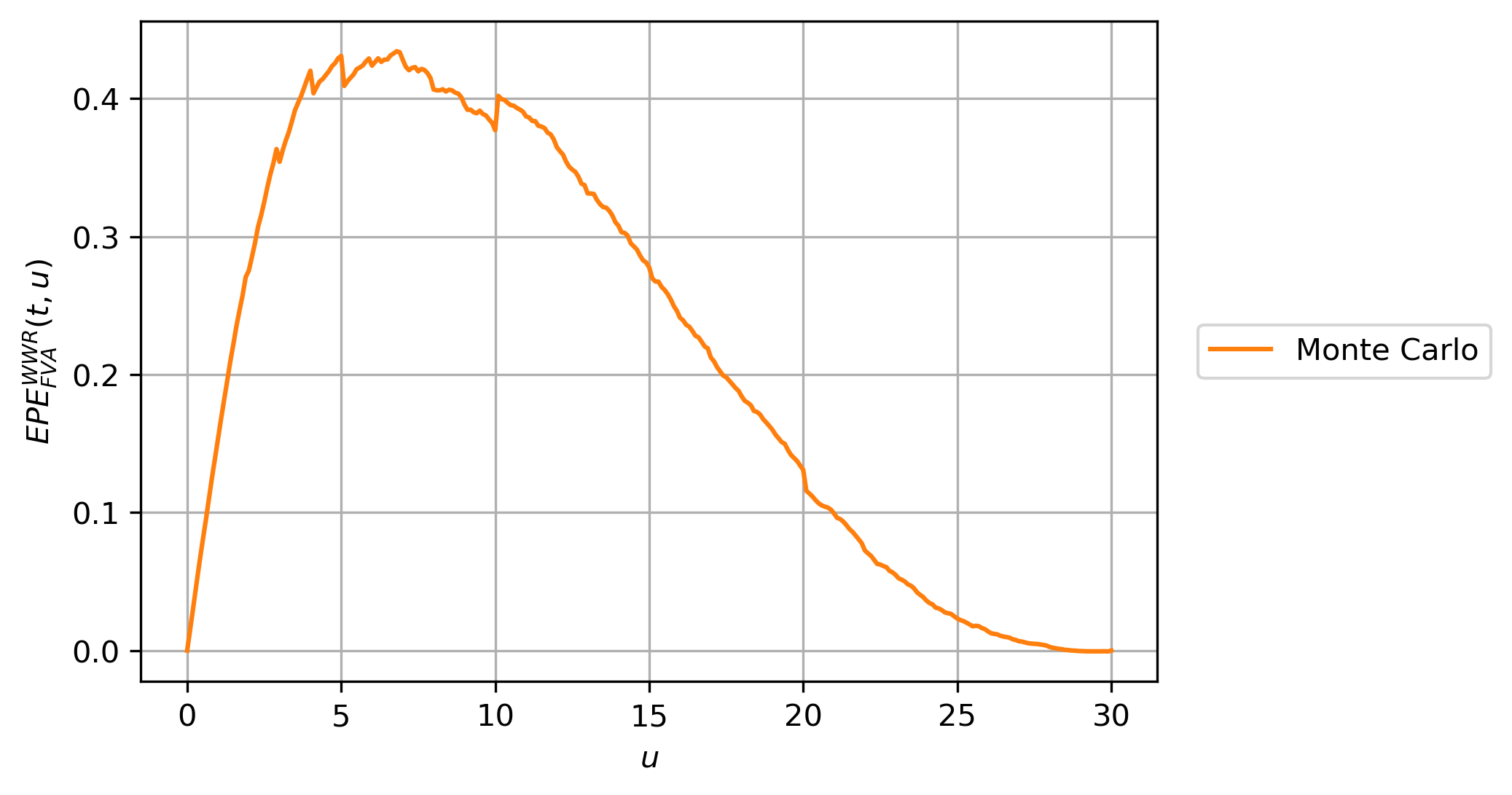}
    \caption{$\EPEFVAWWR{t}{u}$}
    \label{fig:creditITMReceiver31ExclInclWWR}
  \end{subfigure}
  \caption{Stochastic spread, ITM receiver swap, $\default_I$ excl., $\default_C$ incl., $\FVAIndep(t) = 136.5265$.}
  \label{fig:creditITMReceiver31ExclIncl}
\end{figure}

When $\default_I$ is still excluded but $\default_C$ is included in the $\FVA$ definition, both the $\EPEFVA{t}{u}$ and $\EPEFVAWWR{t}{u}$ change significantly in shape and magnitude, see Figure~\ref{fig:creditITMReceiver31ExclIncl}.
This is the result of the different credit curve for $C$, which is different in both shape and magnitude due to the lower credit quality than $I$.
In general, the credit adjustment effect from including $\default_z$, $z\in \{I,C\}$, in the $\FVA$ definition increases for worse credit quality.
In Table~\ref{tab:CreditSpreadParamSet31ReceiverITMFvaResults} it is clear that the inclusion of $\default_C$ results in RWR.
Due to the higher credit volatility for $C$ compared to $I$, this RWR effect is stronger than when only $\default_I$ was included.
This is illustrated by the significantly lower WWR percentage in Table~\ref{tab:CreditSpreadParamSet31ReceiverITMFvaResults}.
Despite the significant reduction in overall WWR (due to the RWR effect from the inclusion of $\default_C$), the WWR effect coming from the funding spread still dominates.

Finally, when including both $\default_I$ and $\default_C$ in the $\FVA$ definition, the plots in Figure~\ref{fig:creditITMReceiver31ExclIncl} do not change significantly in shape, but only get scaled down slightly due to both credit adjustment effects.
As expected, the overall credit adjustment effect in Table~\ref{tab:CreditSpreadParamSet31ReceiverITMFvaResults} is strongest when both $\default_I$ and $\default_C$ are included.
This also holds for the RWR that results from this inclusion, which is demonstrated by the $\FVAWWR(t)$ and `WWR $\%$' numbers, which are the lowest of all results presented so far.

For a deterministic funding spread, the credit adjustment effects are similar as presented so far.
Now, there is no WWR but only RWR coming from the inclusion of $\default_I$ and/or $\default_C$.
In both absolute and relative sense, the RWR numbers are of lower magnitude than those for the stochastic spread.
See for example the results in Table~\ref{tab:IRSpreadParamSet31ReceiverITMFvaResults}, which reports around $6.6\%$ of RWR when both $\default_I$ and $\default_C$ are included.
This should be compared to the change in `WWR $\%$' in Table~\ref{tab:CreditSpreadParamSet31ReceiverITMFvaResults}, where there is around $8.6\%$ of RWR in the stochastic spread case due to the inclusion of both $\default_I$ and $\default_C$.

\begin{table}[h]
  \renewcommand{\arraystretch}{1.1}
  \centering
  \footnotesize
  \begin{tabular}{ll|rrrrr}
     $\default_I$  & $\default_C$   & $\FVAIndep(t)$ & $\FVAWWR(t)$ & $\WWR \%$  \\ \hline
     Incl.& Incl.                   & 123.1260 & -8.1066  & -6.58 \\
  \end{tabular}
  \caption{Deterministic spread, ITM receiver swap.}
  \label{tab:IRSpreadParamSet31ReceiverITMFvaResults}
\end{table}

In summary, both WWR and RWR effects are clearly non-negligible and thus play an important role in $\FVA$ modelling.
For the stochastic spread, the stochastic funding spread results in WWR in this example, and the inclusion of $\default_z$, $z\in \{I,C\}$, generates RWR.
It depends on correlations, credit parameters, IR parameters and product type whether the net result is WWR or RWR.
In case of a deterministic funding spread, there is only RWR from the inclusion of $\default_z$.
All results presented are in line with the conclusions from Section~\ref{sec:wwrRelevance}.

\section{Market data and model parameters} \label{app:extraMarketDataModelParameters}

\subsection{Yield curves} \label{app:yieldCurves}

\begin{table}[H]
  \centering
  \begin{subtable}{.33\linewidth}
    \scriptsize
    \centering
    \begin{tabular}{r|rr}
      $t$ & $\DF(t)$ & $\ZC(t)$ \\ \hline
      0.00 & 1.000000 & 0.00 \\
      0.25 & 0.987578 & 0.05 \\
      0.50 & 0.975310 & 0.05 \\
      0.75 & 0.963194 & 0.05 \\
      1.00 & 0.951229 & 0.05 \\
      1.50 & 0.927743 & 0.05 \\
      2.00 & 0.904837 & 0.05 \\
      2.50 & 0.882497 & 0.05 \\
      3.00 & 0.860708 & 0.05 \\
      4.00 & 0.818731 & 0.05 \\
      5.00 & 0.778801 & 0.05 \\
      6.00 & 0.740818 & 0.05 \\
      7.00 & 0.704688 & 0.05 \\
      8.00 & 0.670320 & 0.05 \\
      9.00 & 0.637628 & 0.05 \\
      10.00 & 0.606531 & 0.05 \\
      12.00 & 0.548812 & 0.05 \\
      15.00 & 0.472367 & 0.05 \\
      20.00 & 0.367879 & 0.05 \\
      25.00 & 0.286505 & 0.05 \\
      30.00 & 0.223130 & 0.05 \\
    \end{tabular}
    \caption{$5\%$ flat ZC curve.}
    \label{tab:curvesYC1}
  \end{subtable}%
  \begin{subtable}{.33\linewidth}
    \scriptsize
    \centering
    \begin{tabular}{r|rr}
      $t$ & $\DF(t)$ & $\ZC(t)$ \\ \hline
      0.00 & 1.000000 & 0.000000 \\
      0.25 & 1.001187 & -0.004744 \\
      0.50 & 1.002448 & -0.004891 \\
      0.75 & 1.003773 & -0.005021 \\
      1.00 & 1.005158 & -0.005145 \\
      1.50 & 1.008088 & -0.005370 \\
      2.00 & 1.011132 & -0.005535 \\
      2.50 & 1.014134 & -0.005614 \\
      3.00 & 1.016990 & -0.005616 \\
      4.00 & 1.022401 & -0.005538 \\
      5.00 & 1.026945 & -0.005318 \\
      6.00 & 1.030583 & -0.005021 \\
      7.00 & 1.033099 & -0.004652 \\
      8.00 & 1.034654 & -0.004258 \\
      9.00 & 1.035117 & -0.003835 \\
      10.00 & 1.034622 & -0.003404 \\
      12.00 & 1.031876 & -0.002615 \\
      15.00 & 1.025681 & -0.001690 \\
      20.00 & 1.021923 & -0.001084 \\
      25.00 & 1.032268 & -0.001270 \\
      30.00 & 1.053926 & -0.001751 \\
    \end{tabular}
    \caption{EUR1D curve.}
    \label{tab:curvesYC2}
  \end{subtable}%
  \caption{Yield curves. $\DF$ denotes discount factor, $\ZC$ denotes zero-coupon.}
  \label{tab:curvesYC}
\end{table}

\subsection{Credit curves} \label{app:creditCurves}

\begin{table}[H]
  \begin{subtable}{.33\linewidth}
    \scriptsize
    \centering
    \begin{tabular}{r|rr}
      $t$ & $\DF(t)$ & $\ZC(t)$ \\ \hline
      0.00 & 1.000000 & 0.000000 \\
      0.50 & 0.998984 & 0.002034 \\
      1.00 & 0.997659 & 0.002343 \\
      2.00 & 0.993528 & 0.003247 \\
      3.00 & 0.987626 & 0.004151 \\
      4.00 & 0.979424 & 0.005198 \\
      5.00 & 0.969391 & 0.006217 \\
      7.00 & 0.946630 & 0.007835 \\
      10.00 & 0.912382 & 0.009170 \\
      15.00 & 0.861670 & 0.009926 \\
      20.00 & 0.813199 & 0.010339 \\
      30.00 & 0.721512 & 0.010880 \\
    \end{tabular}
    \caption{AAA-rating curve.}
    \label{tab:curvesCC1}
  \end{subtable}%
  \begin{subtable}{.33\linewidth}
    \scriptsize
    \centering
    \begin{tabular}{r|rr}
      $t$ & $\DF(t)$ & $\ZC(t)$ \\ \hline
      0.00 & 1.000000 & 0.000000 \\
      0.50 & 0.994676 & 0.010677 \\
      1.00 & 0.988348 & 0.011720 \\
      2.00 & 0.970999 & 0.014715 \\
      3.00 & 0.948562 & 0.017603 \\
      4.00 & 0.920897 & 0.020602 \\
      5.00 & 0.888371 & 0.023673 \\
      7.00 & 0.828067 & 0.026952 \\
      10.00 & 0.745380 & 0.029386 \\
      15.00 & 0.632957 & 0.030490 \\
      20.00 & 0.537460 & 0.031045 \\
      30.00 & 0.386538 & 0.031684 \\
    \end{tabular}
    \caption{BBB-rating  curve.}
    \label{tab:curvesCC2}
  \end{subtable}%
  \begin{subtable}{.33\linewidth}
    \scriptsize
    \centering
    \begin{tabular}{r|rr}
      $t$ & $\DF(t)$ & $\ZC(t)$ \\ \hline
      0.00 & 1.000000 & 0.000000 \\
      0.50 & 0.963312 & 0.074757 \\
      1.00 & 0.919991 & 0.083391 \\
      2.00 & 0.831220 & 0.092431 \\
      3.00 & 0.741957 & 0.099488 \\
      4.00 & 0.657705 & 0.104750 \\
      5.00 & 0.579658 & 0.109064 \\
      7.00 & 0.439022 & 0.117601 \\
      10.00 & 0.296235 & 0.121660 \\
      15.00 & 0.160474 & 0.121975 \\
      20.00 & 0.085857 & 0.122754 \\
      30.00 & 0.023506 & 0.125016 \\
    \end{tabular}
    \caption{B-rating  curve.}
    \label{tab:curvesCC3}
  \end{subtable}
  \caption{Credit curves. $\DF$ denotes discount factor, $\ZC$ denotes zero-coupon.}
  \label{tab:curvesCC}
\end{table}

\subsection{Scenarios}\label{app:scenarios}

\begin{enumerate}
  \item \label{app:scenario1}
    This is the scenario on which the graphs in~\ref{app:additionalResultsModellingAssumptions} are based:
    \begin{itemize}
      \item IR: $x_{\shortRate}(0) = 0.0$, $a_{\shortRate} = 1\e-05$, $\vol_{\shortRate} = 0.00284$, EUR1D yield curve (see Table~\ref{tab:curvesYC2}), ATM $\impliedVolFun{\shortRate} = 0.1$;
      \item Credit for I: $x_I(0) = 0.0016939$, $a_I = 0.05$, $\theta_I = 0.015390$, $\vol_I = 0.02$, $\LGD_I = 0.6$, AAA-rating credit curve (see Table~\ref{tab:curvesCC1}), ATM $\impliedVolFun{I} = 0.07351$;
      \item Credit for C: $x_C(0) = 0.0063774$, $a_C = 0.2$, $\theta_C = 0.035447$, $\vol_C = 0.08$, $\LGD_C = 0.6$, BBB-rating credit curve (see Table~\ref{tab:curvesCC2}), ATM $\impliedVolFun{C} = 0.12090$;
      \item Correlation: $\corr_{\shortRate,I} = -0.35$, $\corr_{\shortRate,C} = -0.5$, $\corr_{I,C} = 0.0$.
    \end{itemize}
  \item \label{app:scenario2}
    This is the base scenario on which most of the scenarios to follow are based:
    \begin{itemize}
      \item IR: $x_{\shortRate}(0) = 0.0$, $a_{\shortRate} = 1\e-05$, $\vol_{\shortRate} = 0.00774$, $5\%$ flat zero-coupon yield curve (see Table~\ref{tab:curvesYC1}), ATM $\impliedVolFun{\shortRate} = 0.1$;
      \item Credit for I: $x_I(0) = 0.0016939$, $a_I = 0.05$, $\theta_I = 0.015390$, $\vol_I = 0.02$, $\LGD_I = 0.6$, AAA-rating credit curve (see Table~\ref{tab:curvesCC1}), ATM $\impliedVolFun{I} = 0.07395$;
      \item Credit for C: $x_C(0) = 0.0016939$, $a_C = 0.05$, $\theta_C = 0.015390$, $\vol_C = 0.02$, $\LGD_C = 0.6$, AAA-rating credit curve (see Table~\ref{tab:curvesCC1}), ATM $\impliedVolFun{C} = 0.07395$;
      \item Correlation: $\corr_{\shortRate,I} = -0.35$, $\corr_{\shortRate,C} = -0.35$, $\corr_{I,C} = 0.0$;
    \end{itemize}
  \item \label{app:scenario3}
    Scenario~\ref{app:scenario2}, but with $\impliedVolFun{\shortRate} = 0.2$ s.t. $\vol_{\shortRate} = 0.01556$.
  \item \label{app:scenario4}
    Scenario~\ref{app:scenario2}, but with $a_{\shortRate} = 0.05$ s.t. $\vol_{\shortRate} = 0.01285$.
  \item \label{app:scenario5}
    Scenario~\ref{app:scenario2}, but with $a_{\shortRate} = 0.05$ and $\impliedVolFun{\shortRate} = 0.2$, s.t.$\vol_{\shortRate} = 0.02578$.
  \item \label{app:scenario6}
    Scenario~\ref{app:scenario2}, but with the EUR1D yield-curve (see Table~\ref{tab:curvesYC2}), s.t. now $\vol_{\shortRate} = 0.00284$ to still get $\impliedVolFun{\shortRate} = 0.1$.
  \item \label{app:scenario7}
    Scenario~\ref{app:scenario2}, but with BBB-rating credit curves for $I$ and $C$ (see Table~\ref{tab:curvesCC2}), s.t. $x_z(0) = 0.0098774$, $\theta_z = 0.041033$ and $\impliedVolFun{z} = 0.05224$, for $z\in\{I,C\}$.
  \item \label{app:scenario8}
    Scenario~\ref{app:scenario2}, but with BBB-rating credit curve for $C$ (see Table~\ref{tab:curvesCC2}), s.t. $x_C(0) = 0.0098774$, $\theta_C = 0.041033$ and $\impliedVolFun{C} = 0.05224$.
  \item \label{app:scenario9}
    Scenario~\ref{app:scenario8}, but with $a_C = 0.2$ and $\vol_C = 0.045$ s.t. we match $\impliedVolFun{C}$ and $\int_0^t b_C(u) \du$ from scenario~\ref{app:scenario8}.
    This results in $x_C(0) = 0.0078774$, $\theta_C = 0.033825$, ATM $\impliedVolFun{C} = 0.07359$.
  \item \label{app:scenario10}
    Scenario~\ref{app:scenario8}, but with B-rating credit curve for $C$ (see Table~\ref{tab:curvesCC3}), $a_C = 0.06$ and $\vol_C = 0.045$ s.t. we match $\impliedVolFun{C}$ from scenario~\ref{app:scenario8} and get as close as possible to $\int_0^t b_C(u) \du$ from scenario~\ref{app:scenario8}.
    This results in $x_C(0) = 0.071957$, $\theta_C = 0.16435$, ATM $\impliedVolFun{C} = 0.07330$.
  \item \label{app:scenario11}
    Scenario~\ref{app:scenario10}, $a_C = 0.02$ and $\vol_C = 0.08$ s.t. the implied volatility is higher.
    In particular, we get $x_C(0) = 0.057657$, $\theta_C = 0.44319$, ATM $\impliedVolFun{C} = 0.13624$.
  \item \label{app:scenario12}
    Scenario~\ref{app:scenario8}, but with $a_I = 0.15$ such that $x_I(0) = 0.0011139$, $\theta_I = 0.012183$, ATM $\impliedVolFun{I} = 0.05871$.
    When increasing $a_I$, we get more curvature in $f^{CIR}(0,t)$, s.t. it is closer to $f^{M}(0,t)$, resulting in a lower $\int_0^t b_I(u) \du$.
    This means that we capture more of the market with the model.
    Furthermore, when increasing $a_I$, $x_I(0)$ and $\theta_I$ and $\impliedVolFun{I}$ go down.
  \item \label{app:scenario13}
    Scenario~\ref{app:scenario8}, but with $\vol_I = 0.04$ s.t. $x_I(0) = 0.0016539$, $\theta_I = 0.016763$, ATM $\impliedVolFun{I} = 0.14155$.
    When increasing $\vol_I$, there are no significant changes in $f^{CIR}(0,t)$, so $\int_0^t b_I(u) \du$ is hardly affected.
    Also, $\impliedVolFun{I}$ scales like $\vol_I$, so a twice as large $\vol_I$ results in a doubling of $\impliedVolFun{I}$.
  \item \label{app:scenario14}
    Scenario~\ref{app:scenario8}, but with $a_I = 0.15$ and $\vol_I = 0.04$ s.t. $x_I(0) = 0.00052392$, $\theta_I = 0.012475$, ATM $\impliedVolFun{I} = 0.11033$.
  \item \label{app:scenario15}
    Scenario~\ref{app:scenario8}, but with $a_C = 0.15$ s.t. $x_C(0) = 0.0088774$, $\theta_C = 0.033506$, ATM $\impliedVolFun{C} = 0.03840$.
  \item \label{app:scenario16}
    Scenario~\ref{app:scenario8}, but with $\vol_C = 0.04$ s.t. $x_C(0) = 0.0097774$, $\theta_C = 0.045113$, ATM $\impliedVolFun{C} = 0.10292$.
  \item \label{app:scenario17}
    Scenario~\ref{app:scenario8}, but with $a_C = 0.15$ and $\vol_C = 0.04$ s.t. $x_C(0) = 0.0087774$, $\theta_C = 0.034305$, ATM $\impliedVolFun{C} = 0.07579$.
  \item \label{app:scenario18}
    Scenario~\ref{app:scenario9}, but with $\corr_{\shortRate,I} = -0.7$.
  \item \label{app:scenario19}
    Scenario~\ref{app:scenario9}, but with $\corr_{\shortRate,C} = -0.7$.
  \item \label{app:scenario20}
    Scenario~\ref{app:scenario9}, but with $\corr_{\shortRate,I} = \corr_{\shortRate,C} = -0.7$.
  \item \label{app:scenario21}
    Scenario~\ref{app:scenario9}, but with $\corr_{\shortRate,I} = \corr_{\shortRate,C} = 0.7$ (opposite sign compared to scenario~\ref{app:scenario20}).
\end{enumerate}

\end{document}